\RequirePackage{fix-cm}
\documentclass[smallextended]{svjour3}       

\smartqed  
\fontfamily{times}
\usepackage{graphicx}
\usepackage{geometry}
\usepackage{amsfonts}
\usepackage{amsmath}
\usepackage{subfigure}

\begin{document}

\title{Factor Analysis of Interval Data}



\author{Paula Cheira         \and
        Paula Brito         \and
        A. Pedro Duarte Silva
}


\institute{Paula Cheira (corresponding author) \at
              ESTG, I.P. Viana do Castelo \& LIAAD-INESC TEC, Univ. Porto, Portugal \\
              Tel.: +351-258-819700 \\
              Fax: +351-258-827636 \\
              \email{paulacheira@estg.ipvc.pt} 
           \and
           Paula Brito \at
              Fac. Economia \& LIAAD-INESC TEC, Univ. Porto, Portugal 
           \and
           A. Pedro Duarte Silva \at
              Cat\'olica Porto Business Scholl \& CEGE, Univ. Cat\'olica Portuguesa/Porto, Portugal             
}

\date{Received: date / Accepted: date}

\maketitle

\begin{abstract}
This paper presents a factor analysis model for symbolic data, focusing on the particular case of interval-valued variables. 
The proposed method describes the correlation structure among the measured interval-valued variables 
in terms of a few underlying, but unobservable, uncorrelated interval-valued variables, called \textit{common factors}. 
Uniform and Triangular distributions are considered within each observed interval. 
We obtain the corresponding sample mean, variance and covariance assuming a general Triangular distribution.

In our proposal, factors are extracted either by Principal Component or by Principal Axis Factoring, performed on the interval-valued variables correlation matrix. 
To estimate the values of the common factors, usually called \textit{factor scores}, 
two approaches are considered, which are inspired in methods for real-valued data: the Bartlett and the Anderson-Rubin methods. 
In both cases, the estimated values are obtained solving an optimization problem that minimizes 
a function of the weighted squared Mallows distance between quantile functions.
Explicit expressions for the quantile function and the squared Mallows distance are derived 
assuming a general Triangular distribution.

The applicability of the method is illustrated using two sets of data: temperature and precipitation in cities of the United States of America between the years 1971 and 2000 and measures of car characteristics of different makes and models. Moreover, the method is evaluated on synthetic data with predefined correlation structures.

\end{abstract}
\section{Introduction}
\label{intro}

When, in 1987, Diday introduced symbolic data, he added a new dimension to data analysis and made us think about data in a new manner. Until then, in multivariate data analysis, data were represented in a data table where for each statistical unit (individual/object) a single value, numerical or categorical, was observed for each variable. However, this structure is unable to represent more complete and complex data, where the information for a statistical unit on each variable cannot be reduced to one single value. Symbolic Data Analysis extends the classical data model with the introduction of new types of statistical units and new types of variables. In this new model the statistical units (entities of interest) may be individuals/objects or classes of individuals/objects, usually called symbolic objects, described by variables, which allow representing explicitly any inherent data variability. We distinguish three new types of variables: interval, multi-valued (numerical and categorical) and distributional (histogram and categorical modal) variables \cite{Billard2006,Bock2000,Brito2014}.

In this paper, we are interested in the analysis of interval data, i.e., where the statistical units are characterized by variables whose realizations are intervals of $\mathbb{R}$. There are numerous situations that give rise to interval data. A natural source of interval data is the aggregation of individual observations described by real values, in groups/classes according to some question of interest, when the databases are too large for direct analysis. Another source of interval data are original symbolic data - examples are descriptions of biological species or technical specifications. \textit{Native} interval data, which occur when describing ranges of variables values, constitute yet another source of interval data - daily stock prices and daily temperatures, are examples of this type of data. Imprecise data, from repeated measures or confidence interval estimation, can also be represented by interval-valued variables.

Since its introduction, the analysis of symbolic data has known a considerable development, becoming one of the new lines of research in Multivariate Statistics and Data Analysis. The present paper introduces a factor analysis model for symbolic data, focusing on the particular case of interval-valued variables. The essential purpose of factor analysis is to explain the covariance and/or correlation structure among the measured variables \cite{Johnson2002,Johnson1998}. In fact, when a large number of variables is measured on each statistical unit, the study of its dependence structure may be of interest. The proposed factor analysis model assumes that there is a smaller set of uncorrelated interval-valued variables - factors - that explain the relations between the interval-valued variables that were actually measured. With the new variables it is expected to get a better understanding of the data being analyzed, moreover, they may be used in future analysis. Two cases are considered for the distribution assumed within each interval: an Uniform distribution and a Triangular distribution \cite{Bertrand2000,Billard2008}.

In our proposal, factors are extracted by Principal Components or by Principal Axis Factoring, performed on the correlation matrix of the interval-valued variables. First and second sample order moments for interval-valued variables have been derived for the Uniform distribution by \cite{Bertrand2000,Billard2003,Billard2008} and for the Symmetric Triangular distribution by \cite{Billard2008}. In this paper, we obtain the formula for the sample mean, variance and covariance assuming a general Triangular distribution.

To estimate the factor scores, two approaches will be considered, which are inspired in methods for real-valued data: the Bartlett and the Anderson-Rubin methods \cite{DiStefano2009}. In  both cases, the estimated values are obtained by solving an optimization problem that uses as criterion to be minimized a function of the weighted squared Mallows distance \cite{Mallows1972} between quantile functions. In the first method the factor scores are highly correlated with their corresponding factor and weakly (or not at all) with other factors. However, the estimated factor scores of different factors may still be correlated. In the second  proposed method, the function to minimize is adapted to ensure that the factor scores are themselves not correlated with each other.

In this work the Mallows distance will be the measure used to evaluate the dissimilarity between distributions. In the last century, several dissimilarity measures between probability distributions were proposed, that Gibbs and Su \cite{Gibbs2002} and Bock and Diday \cite{Bock2000} reviewed and summarized. They present some of the most important metrics on probability measures that are used by statisticians and probabilists, performing a rigorous analysis of their properties and of the relations between them. 
Among all, the Mallows distance\footnote{In many works the Mallows distance is denominated Wasserstein distance. The reason is related to the fact that, historically, this metric has been introduced several times and from different perspectives, and therefore is known under different names \cite{Irpino2015}. However, it was Mallows who introduced this metric in a statistical context, so we will name it henceforth as Mallows distance.} is considered by many researchers in various areas of study, the appropriate measure to evaluate the dissimilarity between probability distributions \cite{Mallows1972,Arroyo2008,Verde2007,Irpino2008}. In the search of the best measure to quantify the error of a forecast, Arroyo \cite{Arroyo2008} studied several divergence measures concluding that only the Mallows and Wasserstein distances are adequate to represent an error measure. According to Arroyo and Mat\'{e} \cite{Arroyo2008,Arroyo2009} these measures have a clear and intuitive interpretation and are the ones that better adjust to the concept of distance as assessed by the human eye. 
Naturally, the Mallows distance was chosen in the Arroyo and Mat\'{e} works \cite{Arroyo2008,Arroyo2009}, on forecasting time series of histogram-valued variables, to measure the error between the observed and forecasted distributions and to calculate the forecasts.
It is noteworthy that the Mallows distance has been successfully applied in several others works in the context of Symbolic Data Analysis. 
Irpino and Verde \cite{Irpino2006} derived the Mallows distance between intervals assuming an Uniform distribution within them. They then used this measure for an agglomerative hierarchical clustering of histogram data \cite{Irpino2006} and as criterion function in a Dynamic Clustering Algorithm (DCA) applied to interval data and histogram data \cite{Irpino2008,Verde2007}.
Irpino and Verde \cite{Irpino2015,Verde2010} and Dias and Brito \cite{Dias2015} also used this distance in a linear regression context. 
Here we deduce the explicit expression of the Mallows distance assuming a general Triangular distribution within each interval.\\

The structure of this paper is as follows: In Section 2 we start by introducing interval-valued variables, review existing interval representations and fix notation. Then we present sample moments, quantile function representations and Mallows distance, assuming an Uniform or a general Triangular distribution within intervals. 
Section 3 presents a factor analysis model for interval-valued variables, where 
factor extraction is done by Principal Components or by Principal Axis Factoring on the correlation matrix between the interval-valued variables. 
Two approaches are considered to estimate the factor scores which are inspired in the Bartlett and the Anderson-Rubin methods. 
Section 4 shows the soundness of our proposal with synthetic data. Section 5 presents applications to one data set of measures of car characteristics of different makes and models and another data set of meteorological data. Section 6 concludes de paper.

\section{Interval-valued variables}
\label{sec:2}

Consider a set of $n$ units $S=\{s_1,\ldots,s_n\}$ under study. An interval-valued variable is defined by an application $Y : S \rightarrow B$ such that for each $s_i \in S$, $Y(s_i)=[l_i, u_i]$, with $l_i \le u_i$, where $B$ is the set of intervals of an underlying set $O \subseteq \mathbb{R}$. Thus the value of an interval-valued variable $Y$ for each $s_i \in S$ is defined by the bounds $l_i$ and $u_i$. Alternatively, the interval $Y(s_i)$ can be represented by its center (midpoint of the interval) $c_i= \displaystyle\frac{l_i+u_i}{2}$ and half-range $r_i= \displaystyle\frac{u_i-l_i}{2}$, then $Y(s_i)=[c_i-r_i, c_i+r_i]$.\\
Let $I$ be an $n \times  p$ matrix representing the values of $p$ interval-valued variables on $S$.
Each unit $s_i \in S$ is represented by a $p$-uple vector of intervals, $I_i=(I_{i1}, ... , I_{ip}), i=1, ... , n,$ with $I_{ij} = [l_{ij}, u_{ij}], j=1,\ldots, p$ (see Table \ref{table1}).

\begin{center}
\begin{table}[h!]
\renewcommand{\arraystretch}{1.5}
\caption{ - Matrix $I$ of interval data.}
\begin{tabular}{|c|ccccc|}
\hline 
 & $Y_{1}$ & \ldots & $Y_{j}$ & \ldots & $Y_{p}$ \\ 
\hline 
$s_{1}$ & \begin{small}$\left[l_{11},u_{11}\right]$\end{small} & \ldots & \begin{small}$\left[l_{1j},u_{1j}\right]$\end{small} & \ldots & \begin{small}$\left[l_{1p},u_{1p}\right]$\end{small} \\ 
\hline 
\vdots & \vdots &  & \vdots &  & \vdots \\  
\hline 
$s_{i}$ & \begin{small}$\left[l_{i1},u_{i1}\right]$\end{small} & \ldots & \begin{small}$\left[l_{ij},u_{ij}\right]$\end{small} & \ldots & \begin{small}$\left[l_{ip},u_{ip}\right]$\end{small} \\  
\hline 
\vdots & \vdots &  & \vdots &  & \vdots \\ 
\hline 
$s_{n}$ & \begin{small}$\left[l_{n1},u_{n1}\right]$\end{small} & \ldots & \begin{small}$\left[l_{nj},u_{nj}\right]$\end{small} & \ldots & \begin{small}$\left[l_{np},u_{np}\right]$\end{small} \\ 
\hline 
\end{tabular}\label{table1}
\end{table}
\end{center}

\textbf{Notation:} 
From now on, in order to simplify, we will denote by $ Y_i $ the value of the interval-valued  variable $Y$ measured on unit $ s_i $ when we are working with only one variable, instead of $ Y(s_i) $. If we measure various variables on the same unit $ s_i $, and we want to mention the values of the variables $ Y_j $ and $ Y_{j'} $, we will write $ Y_{ij} $ and $ Y_{ij'} $, instead of $ Y_j(s_i) $ and $ Y_{j'}(s_i) $.

\subsection{Interval data as quantile functions}
\label{sec2:1}

In this work we will also resort to the representation of the interval $Y_i$ by the respective quantile function (the inverse of the distribution function $\Psi_{Y_i}(y)$), $\Psi_{Y_i}^{-1}(t)$ with $t \in \left[0,1\right]$, assuming a specific distribution within the interval: Uniform distribution or Triangular distribution.\\

If we assume that values within the interval $Y_i=[l_i, u_i]$ follow an Uniform distribution, its representation by the associated quantile function is given by
\begin{equation} \label{QFU1}
\Psi_{Y_i}^{-1}(t)=l_{i}+(u_{i}-l_{i})t, \qquad 0\leq t \leq 1,
\end{equation}
\begin{flushleft}
or
\end{flushleft}
\begin{equation} \label{QFU2}
\Psi_{Y_i}^{-1}(t)=c_{i}+r_{i}(2t-1), \qquad 0\leq t \leq 1,
\end{equation}
\begin{flushleft}
as a function of the center $c_i$ and half-range $r_i$.\\
\end{flushleft}
\vspace{0.25cm}

Assuming a Triangular distribution within interval $Y_i=[l_i, u_i]$, with mode $m_i$, the associated quantile function is as follows :
\begin{equation} \label{QFT1}
\Psi_{Y_i}^{-1}(t)=
  \begin{cases}
    l_{i}+\sqrt{(u_{i}-l_{i})(m_{i}-l_{i})t} , &  \quad 0\leq t \leq \dfrac{m_{i}-l_{i}}{u_{i}-l_{i}}\\
    u_{i}-\sqrt{(u_{i}-l_{i})(u_{i}-m_{i})(1-t)} , &  \quad \dfrac{m_{i}-l_{i}}{u_{i}-l_{i}} < t \leq 1
  \end{cases}
\end{equation}
\begin{flushleft}
or, using the center $c_i$ and half-range $r_i$,
\end{flushleft}
\begin{equation} \label{QFT2}
\Psi_{Y_i}^{-1}(t)=
  \begin{cases}
    c_{i}-r_{i}+\sqrt{2r_{i}(m_{i}-c_{i}+r_{i})t} , &  \quad 0\leq t \leq \dfrac{m_{i}-c_{i}}{2r_{i}}+\dfrac{1}{2}\\
    c_{i}+r_{i}-\sqrt{2r_{i}(c_{i}+r_{i}-m_{i})(1-t)} , &  \quad \dfrac{m_{i}-c_{i}}{2r_{i}}+\dfrac{1}{2} < t \leq 1
  \end{cases}
\end{equation}

In the particular case of the Symmetric Triangular distribution, that is, when $m_i=\dfrac{l_{i}+u_{i}}{2}$, expressions \eqref{QFT1} and \eqref{QFT2} become, respectively,
\begin{equation} \label{QFT3}
\Psi_{Y_i}^{-1}(t)=
  \begin{cases}
    l_{i}+\dfrac{u_{i}-l_{i}}{\sqrt{2}}\sqrt{t} , &  \quad 0\leq t \leq \dfrac{1}{2}\\
    u_{i}-\dfrac{u_{i}-l_{i}}{\sqrt{2}}\sqrt{1-t} , &  \quad \dfrac{1}{2} < t \leq 1
  \end{cases}
\end{equation}
\begin{flushleft}
and
\end{flushleft}
\begin{equation} \label{QFT4}
\Psi_{Y_i}^{-1}(t)=
  \begin{cases}
    c_{i}-r_{i}+r_{i}\sqrt{2t} , &  \quad 0\leq t \leq \dfrac{1}{2}\\
    c_{i}+r_{i}-r_{i}\sqrt{2(1-t)} , &  \quad \dfrac{1}{2} < t \leq 1.
  \end{cases}
\end{equation}\\
It is important to note that if we multiply an interval $Y_i$ by a positive real number $ \lambda $, $ \lambda \Psi_{Y_i}^{-1}(t) $ is the quantile function that represents the resulting interval $\lambda Y_i=[\lambda l_i, \lambda u_i]$, but if $ \lambda $ is a negative real number, the interval $\lambda Y_i=[\lambda u_i, \lambda l_i]$ is represented by the quantile function $ \lambda \Psi_{Y_i}^{-1}(1-t) $. For more details about the behavior of quantile functions, see Dias \cite{Dias2014}.

\subsection{Descriptive statistics}
\label{sec2:2}

All factor analysis models rely on properly defined correlation matrices. In our proposed model, correlations between interval-valued variables are defined as the quotients between covariances and products of standard deviation for interval-valued variables, which depend on the distribution assumed within each interval.  Bertrand and Goupil \cite{Bertrand2000} were the first to propose the univariate statistics for this type of numerical symbolic variables: the mean and variance of a interval variable $Y$, defined on the set of $n$ units $S=\{s_1,\ldots,s_n\}$, correspond to those of a finite mixture of $n$ probability density (or frequency) functions, in which it is assumed that each unit is equally likely to be observed with probability $ \dfrac{1}{n}$. It is well known (see, for instance, \cite{Fruhwirth-Schnatter2006}) that given $n$ variables $Y_i$ with probability density functions $f_i$, $i=1,\ldots,n$, with means $\mu_{i}$ and variances $\sigma_{i}^2$, the variable $Y$ with the finite mixture probability density function $f$:
\begin{equation} \label{Fdp_finmix}
f(y)= \sum_{i=1}^{n} \frac{1}{n} f_i(y) = \frac{1}{n} \sum_{i=1}^{n} f_i(y),
\end{equation}
\begin{flushleft}
\vspace{0,15cm}
has mean and variance, respectively, given by,
\vspace{0,1cm}
\end{flushleft}
\begin{equation} \label{Mean_finmix}
\mu=E \left[Y \right] = \frac{1}{n} \sum_{i=1}^{n} \mu_{i}
\end{equation}
\vspace{0,15cm}
\begin{equation} \label{Var_finmix}
\sigma^2 = E \left[ \left( Y-\mu \right) ^2 \right] 
= \frac{1}{n} \sum_{i=1}^{n} \left( \mu_{i}^2+ \sigma_{i}^2 \right) - \mu^2.
\end{equation}
\\
In the next subsection we present expressions for the symbolic sample mean, the symbolic sample variance and the symbolic sample covariance assuming a Uniform distribution or a Triangular distribution within the intervals.\\

\subsubsection{Uniform distribution}
\label{sec2:2:1}

We now assume an Uniform distribution within each interval $Y(s_i) = I_{i}=[l_{i},u_{i}]$, $i=1,\ldots,n$. 
Under these conditions, Bertrand and Goupil \cite{Bertrand2000} obtained the symbolic sample mean and the symbolic sample variance of the interval-valued variable $Y$, respectively, as
\begin{equation} \label{MeanU_1}
\overline{Y} = \frac{1}{2n}\sum_{i=1}^{n}(l_{i}+u_{i})
\end{equation}
\begin{equation} \label{VarU1_1} 
S_{Y}^2 = \displaystyle \frac{1}{3n}\sum_{i=1}^{n}(l_{i}^2+l_{i}u_{i}+u_{i}^2)-\overline{Y}^2
\end{equation}
\begin{flushleft}
or, expressed in terms of the centers $c_i$ and half-ranges $r_i$ of the interval $I_{i}$,
\end{flushleft}
\begin{equation} \label{MeanU_2}
\overline{Y} = \frac{1}{n}\sum_{i=1}^{n}c_{i} = \frac{1}{n}\sum_{i=1}^{n}\mu_{i}
\end{equation}
\begin{equation} \label{VarU_2} 
S_{Y}^2 = \displaystyle \frac{1}{n}\sum_{i=1}^{n} \Big(\frac{r_{i}^2}{3}+c_{i}^2\Big)-\overline{Y}^2 
= \displaystyle \frac{1}{n}\sum_{i=1}^{n}\sigma_{i}^2 +\displaystyle \frac{1}{n}\sum_{i=1}^{n}\mu_{i}^2-\overline{Y}^2\\,
\end{equation}
\\
obtained from the empirical density function for an interval variable. 
As it can be seen above, the sample variance of the interval-valued variable $Y$ 
is the sum of the average of the (within) variances of observed intervals 
with the variance of the means (or centers) of the intervals.\\

For the symbolic sample covariance three definitions were proposed.\\
Let $Y_{j}$ and $Y_{j'}$ be two interval-valued variables such that for each $s_i \in S=\{s_1,\ldots,s_n\}$, the observed $Y_{ij}$ and $Y_{ij'}$ values, $i=1,\ldots,n$, are uniformly distributed within each interval $I_{ik}=\left[ l_{ik},u_{ik} \right] = \left[ c_{ik}-r_{ik},c_{ik}+r_{ik} \right]$, $k=j,j'$, respectively.\\
The first expression for the sample covariance between two interval-valued variables $Y_{j}$ 
and $Y_{j'}$ was obtained in 2003, by Billard and Diday \cite{Billard2003}, 
from the joint density function, it is denoted Covariance 1 ($Cov_{1}$) :
\begin{equation}  \label{Cov1U_1}
Cov_{1}(Y_{j},Y_{j'}) = \displaystyle \frac{1}{4n}\sum_{i=1}^{n}(l_{ij}+u_{ij})(l_{ij'}+u_{ij'})-\overline{Y}_{j} \overline{Y}_{j'}
\end{equation}
\begin{flushleft}
or, equivalently,
\end{flushleft}
\begin{equation} \label{Cov1U_2}
Cov_{1}(Y_{j},Y_{j'}) =  \displaystyle \frac{1}{n}\sum_{i=1}^{n} c_{ij} c_{ij'}-\overline{Y}_{j} \overline{Y}_{j'}.
\end{equation}\\
Expression \eqref{Cov1U_2} is the classic definition of covariance applied to the centers of the intervals 
$Y_{ij}$ and $Y_{ij'}$. However, it is noted that when one considers the variables $Y_{j} = Y_{j'}$, 
the expression of Covariance 1 does not reduce to the variance in expression \eqref{VarU_2}. 
Furthermore, the resulting Covariance 1 function would not reflect the internal cross-variations between $Y_{j}$ and $Y_{j'}$.\\

In 2006, Billard and Diday \cite{Billard2006} proposed a new expression for the symbolic sample covariance, 
denoted by Covariance 2 ($Cov_{2}$), incorporating more accurately 
both between and within interval variations into the overall covariance,
\begin{equation} \label{Cov2U}
Cov_{2}(Y_{j},Y_{j'}) = \displaystyle \frac{1}{3n}\sum_{i=1}^{n}G_{j}G_{j'}\left[Q_{j}Q_{j'}\right]^{1/2}
\end{equation}
\begin{flushleft}
where, for $k=j,j'$,
\end{flushleft}
\begin{equation*}
Q_{k}=(l_{ik}-\overline{Y}_{k})^2+(l_{ik}-\overline{Y}_{k})(u_{ik}-\overline{Y}_{k})+(u_{ik}-\overline{Y}_{k})^2
\end{equation*}
\begin{equation*}
G_{k}=
  \begin{cases}
    -1  & \text{if} \quad \overline{Y}_{ik} \leq \overline{Y}_{k}\\
    \;1  &  \text{if} \quad \overline{Y}_{ik} > \overline{Y}_{k}
  \end{cases}
\end{equation*}
\begin{flushleft}
and $\overline{Y}_{ik} = \displaystyle \frac{l_{ik}+u_{ik}}{2}$. The $Q_{k}$ and $G_{k}$ 
expressions can be rewritten in terms of the center $c_{ik}$ and half-range $r_{ik}$ of the interval $I_{ik}$ as
\end{flushleft}
\begin{equation*}
Q_{k}=3(c_{ik}-\overline{Y}_{k})^2+r_{ik}^2 \qquad \text{and} \qquad
G_{k}=
  \begin{cases}
    -1  & \text{if} \quad c_{ik} \leq \overline{Y}_{k}\\
    \;1  &  \text{if} \quad c_{ik} > \overline{Y}_{k}
  \end{cases},\quad \text{for} \; k=j,j'. 
\end{equation*}

\vspace{0,25cm}
When the variables $Y_{j} = Y_{j'}$, the expression of Covariance 2 coincides with the expression \eqref{VarU1_1} (or \eqref{VarU_2}) of variance. In 2008 Billard \cite{Billard2008} presented a new formulation, considering a decomposition of the Total Sum of Products (TotalSP), between the variables $Y_{j}$ and $Y_{j'}$, into Within Observations Sum of Products (WithinSP) and Between Observations Sum of Products (BetweenSP), named Covariance 3 ($Cov_{3}$),
\begin{equation} \label{Cov3U_1_1}
Cov_{3}(Y_{j},Y_{j'}) = \displaystyle \frac{1}{n}
\underbrace{\sum\limits_{i=1}^{n}\frac{(u_{ij}-l_{ij})(u_{ij'}-l_{ij'})}{12}}_\text{WithinSP}+\frac{1}{n} 
\underbrace{\sum\limits_{i=1}^{n}\Big(\frac{l_{ij}+u_{ij}}{2}-\overline{Y}_{j}\Big)\Big(\frac{l_{ij'}+u_{ij'}}{2}-\overline{Y}_{j'}\Big)}_\text{BetweenSP}
\end{equation}
\begin{equation} \label{Cov3U_1_2}
\qquad\qquad\qquad\quad = \displaystyle \frac{1}{12n}
\sum\limits_{i=1}^{n}(u_{ij}-l_{ij})(u_{ij'}-l_{ij'})+\frac{1}{4n} 
\sum\limits_{i=1}^{n}(l_{ij}+u_{ij})(l_{ij'}+u_{ij'})-\overline{Y}_{j}\overline{Y}_{j'}
\end{equation}
\begin{flushleft}
or, equivalently,
\end{flushleft}
\begin{equation} \label{Cov3U_2_1}
Cov_{3}(Y_{j},Y_{j'}) = \displaystyle \frac{1}{3n}
\sum\limits_{i=1}^{n}{r_{ij}r_{ij'}}+\frac{1}{n} 
\sum\limits_{i=1}^{n}c_{ij}c_{ij'}-\overline{Y}_{j}\overline{Y}_{j'}
\end{equation}
\begin{equation} \label{Cov3U_2_2}
\qquad\qquad\qquad = \displaystyle \frac{1}{n}
\sum\limits_{i=1}^{n}{\sigma_{ij}\sigma_{ij'}}+\frac{1}{n} 
\sum\limits_{i=1}^{n}\mu_{ij}\mu_{ij'}-\overline{Y}_{j}\overline{Y}_{j'}\\
\end{equation}
\\
Note that the sample Covariance 3 between two interval-valued variables $Y_{j}$ and $Y_{j'}$ is the sum of the average of the product of the standard deviations of the intervals $Y_{ij}$ and $Y_{ij'}$ with the covariance between the means (or centers) of the intervals.
Also noteworthy that expression \eqref{Cov3U_2_1} of covariance reduces to the expression of variance \eqref{VarU_2} when one considers the variables $Y_{j} = Y_{j'}$ and is equivalent to
\begin{equation} \label{Cov3U_2_3}
Cov_{3}(Y_{j},Y_{j'}) = \displaystyle \frac{1}{n}
\sum\limits_{i=1}^{n}{\sigma_{ij}\sigma_{ij'}} + Cov_{1}(Y_{j},Y_{j'})\\
\end{equation}
which shows the relation between the expressions of the covariance 1 and 3.\\

\subsubsection{Triangular distribution}
\label{sec2:2:2}

If a random variable $Y_i$ follows a Triangular distribution within each interval $\left[ l_i, u_i \right]$ with mode $m_{i}$, i.e., $Y_i \sim \mathcal{T}(l_i, u_i, m_i)$ then, the probability density function of $Y_i$ is

\begin{equation} \label{FdpTriangular}
f_i(x)=
  \begin{cases}
    \displaystyle \frac{2(x-l_i)}{(u_i-l_i)(m_i-l_i)}  & \text{for} \quad l_i \leq x \leq m_i\\
     \displaystyle \frac{2(u_i-x)}{(u_i-l_i)(u_i-m_i)}  & \text{for} \quad m_i < x \leq u_i\\    
    \qquad \quad 0  & \text{for any other case}
  \end{cases}
\end{equation}
\begin{flushleft}
with mean and variance
\end{flushleft}
\begin{equation} \label{MeanVarT} 
\mu_{i} = E(Y_{i}) = \frac{l_{i}+u_{i}+m_{i}}{3} = \frac{2c_{i}+m_{i}}{3}
\end{equation}
\begin{equation} \label{VarVarT} 
\sigma_{i}^2 = Var(Y_{i}) = \frac{l_{i}^2+u_{i}^2+m_{i}^2-l_{i}u_{i}-l_{i}m_{i}-u_{i}m_{i}}{18} = \frac{\left(c_{i}-m_{i}\right)^2}{18} + \frac{r_{i}^2}{6},
\end{equation}\\
where $c_{i}$ and $r_{i}$ are, respectively, the center and the half-range of the interval $\left[l_i, u_i\right]$.\\
\\ 
Under the assumption that the observed $Y_i$ values, for each $s_i \in S=\{s_1,\ldots,s_n\}$, follow a Triangular distribution within each interval $[l_i, u_i]$ with mode $m_{i}$, $i=1,\ldots,n$, the symbolic sample mean and the symbolic sample variance of the interval variable $Y$ are given, respectively, by,  
\begin{equation} \label{MeanT_1}
\overline{Y}= \displaystyle \frac{1}{3n}\sum_{i=1}^{n} \left( l_{i}+u_{i}+m_{i} \right) 
\end{equation}
\begin{equation} \label{VarT_1}
S_{Y}^2=\displaystyle \frac{1}{6n}\sum_{i=1}^{n} \left( l_{i}^2+u_{i}^2+m_{i}^2+l_{i}u_{i}+l_{i}m_{i}+u_{i}m_{i} \right) -\overline{Y}^2,
\end{equation}

\begin{flushleft}
obtained from the empirical density function, following the same line of reasoning of Bertrand and Goupil \cite{Bertrand2000}, in determining expressions \eqref{MeanU_1} and \eqref{VarU1_1}. Expressions \eqref{MeanT_1} and \eqref{VarT_1} may be rewritten as
\end{flushleft}
\begin{equation} \label{MeanT_2}
\overline{Y}= \displaystyle \frac{1}{3n}\sum_{i=1}^{n} \left( 2c_{i}+m_{i}\right) = \frac{1}{n}\sum_{i=1}^{n}\mu_{i}
\end{equation}
\begin{equation} \label{VarT_2}
S_{Y}^2=\displaystyle \frac{1}{6n} \sum_{i=1}^{n} \left( 3c_{i}^2+2c_{i}m_{i}+m_{i}^2+r_{i}^2 \right) -\overline{Y}^2 
= \displaystyle \frac{1}{n}\sum_{i=1}^{n}\sigma_{i}^2 +\displaystyle \frac{1}{n}\sum_{i=1}^{n}\mu_{i}^2-\overline{Y}^2,
\end{equation} 
\\
when expressed in terms of the center $c_i$ and half-range $r_i$ of the interval $[l_i, u_i]$. Similarly to what occurs with the Uniform distribution, also the sample variance of the interval-valued variable $Y$ with Triangular distribution can be written as the sum of the average of the variances of the intervals with the variance of the centers of the intervals.\\
\\
Following the same reasoning presented for the Uniform distribution \cite{Billard2003,Billard2006,Billard2008} we can also get three expressions for the symbolic sample covariance between two interval variables.\\
Suppose $Y_{j}$ and $Y_{j'}$ are two interval-valued variables, such that for each $s_i \in S=\{s_1,\ldots,s_n\}$, the observed $Y_{ij}$ and $Y_{ij'}$ values follow a Triangular distribution within each interval $I_{ik}=\left[ l_{ik},u_{ik} \right] = \left[ c_{ik}-r_{ik},c_{ik}+r_{ik} \right]$ with mode $m_{ik}$, $k=j,j'$, respectively, $i=1,\ldots,n$.\\
From the joint density function an expression for the covariance was obtained, which we will denote by Covariance 1 ($Cov_{1}$), 
\begin{equation} \label{Cov1T_1}
Cov_{1}(Y_{j},Y_{j'}) = \displaystyle \frac{1}{9n}\sum_{i=1}^{n}
(l_{ij}+u_{ij}+m_{ij})(l_{ij'}+u_{ij'}+m_{ij'})-\overline{Y}_{j}.\overline{Y}_{j'}
\end{equation}
\begin{flushleft}
or, equivalently,
\end{flushleft}
\begin{equation} \label{Cov1T_2}
Cov_{1}(Y_{j},Y_{j'}) =  \displaystyle \frac{1}{9n}\sum_{i=1}^{n} (2c_{ij}+m_{ij})(2c_{ij'}+m_{ij'})-\overline{Y}_{j}.\overline{Y}_{j'}.
\end{equation}
\vspace{0,25cm}
Incorporating more accurately both between and within interval variations into the overall covariance, a new expression can be defined for symbolic sample covariance, denoted by Covariance 2 ($Cov_{2}$), as follows
\begin{equation} \label{Cov2T}
Cov_{2}(Y_{j},Y_{j'}) = \displaystyle \frac{1}{6n}\sum_{i=1}^{n}G_{j}G_{j'}\left[Q_{j},Q_{j'}\right]^{1/2}
\end{equation}
\begin{flushleft}
with,
\end{flushleft} 
\begin{center}
$Q_{k}=(l_{ik}-\overline{Y}_{k})^2+(u_{ik}-\overline{Y}_{k})^2+(m_{ik}-\overline{Y}_{k})^2+(l_{ik}-\overline{Y}_{k})(u_{ik}-\overline{Y}_{k})
+$\\
\vspace{0,2cm}
$(l_{ik}-\overline{Y}_{k})(m_{ik}-\overline{Y}_{k})
+(u_{ik}-\overline{Y}_{k})(m_{ik}-\overline{Y}_{k})$,\\
\end{center}
\begin{equation*}
G_{k}=
  \begin{cases}
    -1  & \text{if} \quad \overline{Y}_{ik} \leq \overline{Y}_{k}\\
    \;1  &  \text{if} \quad \overline{Y}_{ik} > \overline{Y}_{k}
  \end{cases}
\end{equation*}\\
and $\overline{Y}_{ik}= \displaystyle \frac{l_{ik}+u_{ik}+m_{ik}}{3}$, for $k=j,j'$. 
$Q_{k}$ can be rewritten in terms of the center $c_{ik}$ and half-range $r_{ik}$ of the interval $I_{ik}$ as
\begin{equation*}
Q_{k}=3(c_{ik}-\overline{Y}_{k})^2+2(c_{ik}-\overline{Y}_{k})(m_{ik}-\overline{Y}_{k})+(m_{ik}-\overline{Y}_{k})^2+r_{ik}^2.
\end{equation*}

Considering a decomposition of the Total Sum of Products (TotalSP), between the variables $Y_{j}$ and $Y_{j'}$, into Within Observations Sum of Products (WithinSP) and Between Observations Sum of Products (BetweenSP), as suggested by Billard \cite{Billard2008}, we obtain a new expression for the covariance named Covariance 3, ($Cov_{3}$):
\begin{eqnarray*} \label{Cov3T_1}
Cov_{3}(Y_{j},Y_{j'}) = \displaystyle \frac{1}{n}
\underbrace{\sum\limits_{i=1}^{n}\sqrt{\dfrac{W_{ij}}{18}\dfrac{W_{ij'}}{18}}}_\text{WithinSP}
+\frac{1}{n} \underbrace{\sum\limits_{i=1}^{n}\Big(\frac{l_{ij}+u_{ij}+m_{ij}}{3}-\overline{Y}_{j}\Big)\Big(\frac{l_{ij'}+u_{ij'}+m_{ij'}}{3}-\overline{Y}_{j'}\Big)}_\text{BetweenSP}
\end{eqnarray*} 
\begin{eqnarray} \label{Cov3T_2}
\qquad\qquad\;= \displaystyle \frac{1}{18n}\sum\limits_{i=1}^{n}\sqrt{W_{ij}W_{ij'}}
+\frac{1}{9n}\sum\limits_{i=1}^{n}(l_{ij}+u_{ij}+m_{ij})(l_{ij'}+u_{ij'}+m_{ij'})-\overline{Y}_{j}\overline{Y}_{j'}
\end{eqnarray} 

\begin{flushleft}
where, \\
\begin{equation}
W_{ik}=l_{ik}^2+u_{ik}^2+m_{ik}^2-l_{ik}u_{ik}-l_{ik}m_{ik}-u_{ik}m_{ik},\quad \text{for} \; k=j,j' .
\end{equation}
\end{flushleft}
This last expression of covariance may be written in terms of the center $c_{ik}$ and half-range $r_{ik}$ of the interval $I_{ik}$ as:
\begin{eqnarray} \label{Cov3T_3_1}
Cov_{3}(Y_{j},Y_{j'}) = \displaystyle \frac{1}{18n}\sum\limits_{i=1}^{n}\sqrt{\Big( \left( c_{ij}-m_{ij} \right)^2 + 3r_{ij}^2 \Big)\Big( \left( c_{ij'}-m_{ij'} \right)^2 + 3r_{ij'}^2 \Big)}\nonumber\\
+\frac{1}{9n}\sum\limits_{i=1}^{n}(2c_{ij}+m_{ij})(2c_{ij'}+m_{ij'})-\overline{Y}_{j}\overline{Y}_{j'}
\end{eqnarray}
\begin{equation} \label{Cov3T_3_2}
= \displaystyle \frac{1}{n}
\sum\limits_{i=1}^{n}{\sigma_{ij}\sigma_{ij'}}+\frac{1}{n} 
\sum\limits_{i=1}^{n}\mu_{ij}\mu_{ij'}-\overline{Y}_{j}\overline{Y}_{j'} \quad \quad \quad \quad
\end{equation}\\
When the variables $Y_{j} = Y_{j'}$, the expressions of Covariance 2 and 3 coincide 
with the expression \eqref{VarT_1} (or \eqref{VarT_2}) of variance, but this is not the case for the  expression of Covariance 1. 
As previously, the sample Covariance 3 between two interval-valued variables 
$Y_{j}$ and $Y_{j'}$ is the sum of the average of the product of the standard 
deviations of the intervals $Y_{ij}$ and $Y_{ij'}$ with the covariance between the means of the intervals. 
In addition, it is related with  Covariance 1 according to the following expression,
\begin{equation} \label{Cov3T_3_3}
Cov_{3}(Y_{j},Y_{j'}) = \displaystyle \frac{1}{n}
\sum\limits_{i=1}^{n}{\sigma_{ij}\sigma_{ij'}} + Cov_{1}(Y_{j},Y_{j'})
\end{equation}
\\
When the observed $Y_i$ values, for each $s_i \in S=\{s_1,\ldots,s_n\}$, follow a Triangular distribution within the interval $[l_i, u_i]$ and the mode $m_{i}$ coincides with the center of the interval, i.e. 
$Y_i \sim \mathcal{T}(l_i, u_i, c_i)$, it is said that it follows a Symmetric Triangular distribution. 
In this case, the expressions presented above become much simpler. The mean and variance of $Y_i$  are given by \cite{Billard2008}:
\begin{equation} \label{MeanVarTS} 
\mu_{i} = E(Y_{i}) = \frac{l_{i}+u_{i}}{2} = c_{i}
\end{equation}
\begin{equation} \label{VarVarTS} 
\sigma_{i}^2 = Var(Y_{i}) = \frac{(u_{i}-l_{i})^2}{24} = \frac{r_{i}^2}{6}.\\
\end{equation}
\\
The sample mean and variance of the interval variable Y becomes \cite{Billard2008}:
\begin{equation*}
\overline{Y}= \displaystyle \frac{1}{2n}\sum_{i=1}^{n} (l_{i}+u_{i})
\end{equation*}
\begin{equation} \label{MeanTS}
\qquad\quad = \displaystyle \frac{1}{n}\sum_{i=1}^{n} c_{i}=\displaystyle \frac{1}{n}\sum_{i=1}^{n} \mu_{i}
\end{equation}
\begin{equation*}
S_{Y}^2=\displaystyle \frac{1}{24n}\sum_{i=1}^{n} (7l_{i}^2+10l_{i}u_{i}+7u_{i}^2) -\overline{Y}^2
\end{equation*}
\begin{equation} \label{VarTS}
\quad\qquad\qquad\qquad\qquad = \displaystyle \frac{1}{n}\sum_{i=1}^{n} \Big(\frac{r_{i}^2}{6}+c_{i}^2\Big) -\overline{Y}^2
= \displaystyle \frac{1}{n}\sum_{i=1}^{n}\sigma_{i}^2 +\displaystyle \frac{1}{n}\sum_{i=1}^{n}\mu_{i}^2-\overline{Y}^2,
\end{equation}

\begin{flushleft}
and the expressions of covariance between $Y_j$ and $Y_{j'}$ interval-valued variables as follows,
\end{flushleft}
\begin{equation} \label{Cov1TS}
Cov_{1}(Y_{j},Y_{j'}) = \displaystyle \frac{1}{4n}\sum_{i=1}^{n}
(l_{ij}+u_{ij})(l_{ij'}+u_{ij'})-\overline{Y}_{j}.\overline{Y}_{j'}
=  \displaystyle \frac{1}{n}\sum_{i=1}^{n} c_{ij}c_{ij'}-\overline{Y}_{j}.\overline{Y}_{j'}.
\end{equation}
\begin{equation} \label{Cov2TS}
Cov_{2}(Y_{j},Y_{j'}) = \displaystyle \frac{1}{6n}\sum_{i=1}^{n}G_{j}G_{j'}\left[Q_{j},Q_{j'}\right]^{1/2}
\end{equation}
\begin{flushleft}
with,
\end{flushleft} 
\begin{center}
$Q_{k}= \displaystyle \frac{7}{4}(l_{ik}-\overline{Y}_{k})^2+\frac{7}{4}(u_{ik}-\overline{Y}_{k})^2+
\frac{5}{2}(l_{ik}-\overline{Y}_{k})(u_{ik}-\overline{Y}_{k})=6(c_{ik}-\overline{Y}_{k})^2+r_{ik}^2$\\
\end{center}
\begin{equation*}
G_{k}=
  \begin{cases}
    -1  & \text{if} \quad \overline{Y}_{ik} \leq \overline{Y}_{k}\\
    \;1  &  \text{if} \quad \overline{Y}_{ik} > \overline{Y}_{k}
  \end{cases}
\end{equation*}\\
and $\overline{Y}_{ik}= \displaystyle \frac{l_{ik}+u_{ik}}{2}=c_{ik}$, for $k=j,j'$.\\
\begin{equation} \label{Cov3TS_1}
Cov_{3}(Y_{j},Y_{j'}) = \displaystyle \frac{1}{24n}
\sum\limits_{i=1}^{n}(u_{ij}-l_{ij})(u_{ij'}-l_{ij'})+\frac{1}{4n} 
\sum\limits_{i=1}^{n}(l_{ij}+u_{ij})(l_{ij'}+u_{ij'})-\overline{Y}_{j}\overline{Y}_{j'}
\end{equation}
\begin{flushleft}
or
\end{flushleft} 
\begin{equation} \label{Cov3TS_2}
Cov_{3}(Y_{j},Y_{j'}) = \displaystyle \frac{1}{6n}
\sum\limits_{i=1}^{n}{r_{ij}r_{ij'}}+\frac{1}{n} 
\sum\limits_{i=1}^{n}c_{ij}c_{ij'}-\overline{Y}_{j}\overline{Y}_{j'}.
\end{equation}
\begin{equation} \label{Cov3TS_3}
\qquad \qquad \qquad = \displaystyle \frac{1}{n}
\sum\limits_{i=1}^{n}{\sigma_{ij}\sigma_{ij'}}+\frac{1}{n} 
\sum\limits_{i=1}^{n}\mu_{ij}\mu_{ij'}-\overline{Y}_{j}\overline{Y}_{j'} 
\end{equation}\\
It should be noted that expression \eqref{Cov3TS_1} is equivalent to that presented in 2008 by Billard \cite{Billard2008}.

\subsection{Measuring distances by quantile functions}
\label{sec2:3}

The goal of this work is to propose a factor model, in which the observed interval 
variables are written as linear combinations of a few unobservable variables, 
the \textit{common factors}, also defined as ranges of values. 
Although in factor analysis, the interest is usually focused on the parameters of the factor model, 
the estimated values of the common factors, the \textit{factor scores}, are also often required. \\
In classical factor analysis, one of the methods of estimation of factor scores was suggested by Bartlett, 
and is known as the Weighted Least Squares method. Bartlett \cite{Bartlett1937} proposed to choose as estimates of factor scores 
those that minimize the sum of squared errors, weighted by the reciprocal of their variances. 
In this case, the error is calculated as the difference between two real numbers, 
the observed variable value and the linear combination of common factors, 
and will be the smaller the closer these numbers are. \\
In the factor model that we propose, both the observed variables values and 
the result of the linear combination of the common factors are intervals and 
therefore it would be natural that the error evaluation was measured by 
the difference between those intervals. However, the difference between intervals 
is not the appropriate measure to calculate the similarity between them \cite{Moore2009}. 
In fact, the difference between two equal intervals is not equal to the null interval [0,0], 
but in a interval with center zero and symmetrical bounds. Moreover the difference between 
any two ranges of values, non-degenerate, never results in the null interval. 
This happens because in interval arithmetic the resulting interval, 
from any of the four basic arithmetic operations between intervals, 
includes all results that are possible to be obtained with all pairs of numbers, 
one from each of the two intervals, respectively (with the only restriction 
that zero can not belong to the second interval if the operation is the division). 
That is, if we consider two intervals $X$ and $Y$, $X \odot Y = \lbrace x \odot y : x\in X, y\in Y\rbrace$, 
where $\odot$ represents any of the four arithmetic operations. 
In the particular case of the arithmetic \textit{difference}, we obtain
$$X - Y = \lbrace x - y : x\in X, y\in Y\rbrace$$
\begin{flushleft}
or,
\end{flushleft} 
$$X - Y = \left[ l_{x}-u_{y},u_{x}-l_{y}\right],$$
\begin{flushleft}
for $X=\left[ l_{x},u_{x}\right] $ and $Y=\left[ l_{y},u_{y}\right]$.
\end{flushleft}

Therefore, at this step, it is necessary to select an appropriate measure of similarity between intervals, 
since it is inappropriate to apply arithmetic operations.  
Two measures considered to be good choice to study the dissimilarity between data with variability, in particular data with interval-valued variables or histogram-valued variables, are the Mallows and the Wassertein distances \cite{Arroyo2008,Verde2007,Irpino2008}. 
As stated above, the Mallows and Wasserstein distances are appropriate to represent an error measure.
Furthermore, as distances they present interesting properties: 
they are positive definite measures, symmetric, and satisfy the triangular inequality. 
Both the Mallows distance and the Wasserstein distance are defined in terms of quantile functions, 
and the further apart these functions are, the greater the distance between them.\\
Following is the definition of Mallows and Wasserstein distances:

\begin{definition}\label{defMallows}
If $Y_{j}$ and $Y_{j'}$ are interval-valued variables represented, 
respectively, by their quantile functions $\Psi_{Y_{ij}}^{-1}$ and 
$\Psi_{Y_{ij'}}^{-1}$ for an observation $s_i$, then 
the \textbf{Wasserstein distance} between intervals $Y_{ij}$ and $Y_{ij'}$ is defined by:
\begin{equation}
D_W(\Psi_{Y_{ij}}^{-1}(t),\Psi_{Y_{ij'}}^{-1}(t))=\int_{0}^{1} | \Psi_{Y_{ij}}^{-1}(t)-\Psi_{Y_{ij'}}^{-1}(t) | dt,
\end{equation}
\begin{flushleft}
and the \textbf{squared Mallows distance},
\end{flushleft}
\begin{equation}  \label{SquareMallowsDist}
D^{2}_M(\Psi_{Y_{ij}}^{-1}(t),\Psi_{Y_{ij'}}^{-1}(t))=\int_{0}^{1}(\Psi_{Y_{ij}}^{-1}(t)-\Psi_{Y_{ij'}}^{-1}(t))^2dt.
\end{equation}
\end{definition}

\vspace{0.2cm}
According to Arroyo \cite{Arroyo2008}, one can establish a parallelism between the Wasserstein and the Manhattan distances, 
and between the Mallows and the Euclidean distances. In fact, taking into account the general expression 
\begin{equation} \label{DistGeral}
D(\Psi_{Y_{ij}}^{-1}(t),\Psi_{Y_{ij'}}^{-1}(t))= \Big(\int_{0}^{1}(\Psi_{Y_{ij}}^{-1}(t)-\Psi_{Y_{ij'}}^{-1}(t))^pdt \Big)^{\frac{1}{p}}
\end{equation}

\begin{flushleft}
the Wasserstein and Mallows distances are obtained when \textit{p} = 1 and \textit{p} = 2, respectively.
\end{flushleft}
Also notice that when $Y_{ij}$ and $Y_{ij'}$ are degenerate intervals, 
and considering that expression \eqref{DistGeral} is similar to the Minkowski metric, 
we then obtain the Manhattan distance for the particular case of \textit{p} = 1, and the Euclidean distance for \textit{p} = 2.\\

Irpino and Verde \cite{Irpino2006} have rewritten expression \eqref{SquareMallowsDist} 
assuming the Uniform distribution within each interval,
using the centres and half-ranges:

\begin{proposition}\label{defMallowsUnif}
\cite{Irpino2006} If $Y_{j}$ and $Y_{j'}$ are interval-valued variables represented, 
respectively, by their quantile functions $\Psi_{Y_{ij}}^{-1}$ and $\Psi_{Y_{ij'}}^{-1}$ 
for an observation $s_i$, and the \textbf{Uniform distribution} is assumed
within the intervals $ I_{ik} $, $ k=j,j' $, respectively, 
then the \textbf{squared Mallows distance} between intervals is given by:
\begin{equation} \label{DMUnif}
D^{2}_M(\Psi_{Y_{ij}}^{-1}(t),\Psi_{Y_{ij'}}^{-1}(t))=\displaystyle (c_{ij}-c_{ij'})^2+\dfrac{1}{3}(r_{ij}-r_{ij'})^2
\end{equation}
where, $c_{ij}$, $c_{ij'}$ and $r_{ij}$, $r_{ij'}$ are, respectively, 
the centers and the half-ranges of the observed $Y_{ij}$ and $Y_{ij'}$ intervals.\\
\end{proposition}

Next we deduce the expression of the (square) Mallows distance for the case where 
the Triangular distribution is assumed within the observed intervals.

\begin{proposition}\label{defMallowsTriang}
If $Y_{j}$ and $Y_{j'}$ are interval-valued variables represented by their quantile functions $\Psi_{Y_{ij}}^{-1}$ and $\Psi_{Y_{ij'}}^{-1}$ 
for an observation $s_i$, and the \textbf{Triangular distribution} is assumed  
within the intervals $I_{ik} $, with mode $ m_{ik} $, $ k=j,j' $, then the \textbf{square of the Mallows distance} between intervals is given by, respectively:\\
\textsc{(i)} \textbf{non-degenerated intervals}, i.e., intervals with non-zero half-ranges:

\begin{flushleft}
\hspace{0.2cm} if $ \dfrac{m_j-c_j}{2r_j} \leq \dfrac{m_{ij'}-c_{ij'}}{2r_{ij'}} $,
\end{flushleft}
\begin{equation*} 
\; \; \; \; \; \; D^{2}_M(\Psi_{Y_{ij}}^{-1}(t),\Psi_{Y_{ij'}}^{-1}(t))=
\displaystyle (c_{ij}-c_{ij'})^2+\dfrac{1}{6}(r_{ij}-r_{ij'})^2+\dfrac{1}{6}(m_{ij}-c_{ij})^2+\dfrac{1}{6}(m_{ij'}-c_{ij'})^2 \qquad\qquad\qquad\qquad\qquad
\end{equation*}
\begin{equation*} 
\quad-\dfrac{5}{3}r_{ij}r_{ij'}+\dfrac{2}{3}(m_{ij}-c_{ij})(c_{ij}-c_{ij'}+r_{ij'})-\dfrac{2}{3}(m_{ij'}-c_{ij'})(c_{ij}-c_{ij'}+r_{ij})\qquad\qquad\qquad
\end{equation*}
\begin{equation*} 
+\dfrac{1}{6}\sqrt{r_jr{ij}_{ij'}(m_{ij}-c_{ij}+r_{ij})(m_{ij'}-c_{ij'}+r_{ij'})}\Big(5-\dfrac{m_{ij}-c_{ij}}{r_{ij}}\Big)\qquad\qquad\qquad\qquad\quad\;
\end{equation*}
\begin{equation*} 
+\dfrac{1}{6}\sqrt{r_{ij}r_{ij'}(c_{ij}+r_{ij}-m_{ij})(c_{ij'}+r_{ij'}-m_{ij'})}\Big(5+\dfrac{m_{ij'}-c_{ij'}}{r_{ij'}}\Big) \qquad\qquad\qquad\qquad\quad
\end{equation*}
\begin{equation}  \label{DMTGeral-ramo1}
\quad\quad\;+\dfrac{1}{2}\sqrt{r_{ij}r_{ij'}(c_{ij}+r_{ij}-m_{ij})(m_{ij'}-c_{ij'}+r_{ij'})}\Big(arcsin\dfrac{m_{ij'}-c_{ij'}}{r_{ij'}}-arcsin\dfrac{m_{ij}-c_{ij}}{r_{ij}}\Big)
\end{equation}
\begin{flushleft}
	
\hspace{0.2cm} if $ \dfrac{m_{ij}-c_{ij}}{2r_{ij}} > \dfrac{m_{ij'}-c_{ij'}}{2r_{ij'}} $,
\end{flushleft}
\begin{equation*} 
\; \; \; \; \; \; D^{2}_M(\Psi_{Y_{ij}}^{-1}(t),\Psi_{Y_{ij'}}^{-1}(t))=
\displaystyle (c_{ij}-c_{ij'})^2+\dfrac{1}{6}(r_{ij}-r_{ij'})^2+\dfrac{1}{6}(m_{ij}-c_{ij})^2+\dfrac{1}{6}(m_{ij'}-c_{ij'})^2\qquad\qquad\qquad\qquad\qquad
\end{equation*}
\begin{equation*} 
\quad-\dfrac{5}{3}r_{ij}r_{ij'}+\dfrac{2}{3}(m_{ij}-c_{ij})(c_{ij}-c_{ij'}-r_{ij'})-\dfrac{2}{3}(m_{ij'}-c_{ij'})(c_{ij}-c_{ij'}-r_{ij}) \qquad\qquad\qquad
\end{equation*}
\begin{equation*} 
+\dfrac{1}{6}\sqrt{r_{ij}r_{ij'}(m_{ij}-c_{ij}+r_{ij})(m_{ij'}-c_{ij'}+r_{ij'})}\Big(5-\dfrac{m_{ij'}-c_{ij'}}{r_{ij'}}\Big)\qquad\qquad\qquad\qquad\quad\;
\end{equation*}
\begin{equation*} 
+\dfrac{1}{6}\sqrt{r_{ij}r_{ij'}(c_{ij}+r_{ij}-m_{ij})(c_{ij'}+r_{ij'}-m_{ij'})}\Big(5+\dfrac{m_{ij}-c_{ij}}{r_{ij}}\Big) \qquad\qquad\qquad\qquad\quad\quad
\end{equation*}
\begin{equation} \label{DMTGeral-ramo2}
\quad\quad\;+\dfrac{1}{2}\sqrt{r_{ij}r_{ij'}(c_{ij'}+r_{ij'}-m_{ij'})(m_{ij}-c_{ij}+r_{ij})}\Big(arcsin\dfrac{m_{ij}-c_{ij}}{r_{ij}}-arcsin\dfrac{m_{ij'}-c_{ij'}}{r_{ij'}}\Big)
\end{equation}
\hspace{0.15cm} where, $c_{ij}$,$c_{ij'}$ and $r_{ij}$, $r_{ij'}$ are, respectively, the centers and the half-ranges of the observed $Y_{ij}$ and $Y_{ij'}$ intervals;\\

\textsc{(ii)} only one \textbf{non-degenerate interval}: for instance $ I_{ij'}=[c_{ij'},c_{ij'}] $
\begin{equation*} 
\; \; \; D^{2}_M(\Psi_{Y_{ij}}^{-1}(t),\Psi_{Y_{ij'}}^{-1}(t))=\displaystyle (c_{ij}-c_{ij'})^2-\dfrac{4}{3}(m_{ij}-c_{ij})^2-\dfrac{1}{3}r_{ij}^2+\dfrac{2}{3}(m_{ij}-c_{ij})(c_{ij}-c_{ij'}) \qquad
\end{equation*}
\begin{equation} 
+\dfrac{(m_{ij}-c_{ij}+r_{ij})^3}{4r_{ij}}+\dfrac{(c_{ij}+r_{ij}-m_{ij})^3}{4r_{ij}}
\end{equation}\\

\textsc{(iii)} \textbf{degenerated intervals}: $ I_{ij}=[c_{ij},c_{ij}] $ and $ I_{ij'}=[c_{ij'},c_{ij'}] $
\begin{equation} 
D^{2}_M(\Psi_{Y_{ij}}^{-1}(t),\Psi_{Y_{ij'}}^{-1}(t))=\displaystyle (c_{ij}-c_{ij'})^2
\end{equation}
\vspace{0,01cm}
\hspace{0.3cm}
or equivalently
\begin{equation} \label{DMReais}
D_M(\Psi_{Y_{ij}}^{-1}(t),\Psi_{Y_{ij'}}^{-1}(t))=|c_{ij}-c_{ij'}|.
\end{equation}
\end{proposition}

\vspace{0.4cm}

Expression \eqref{DMReais} is the Euclidean distance between two real numbers, 
$ c_{ij} $ and  $ c_{ij'} $, as expected. 
Note that we obtain exactly the same result if we consider null 
$r_{ij} $ and  $r_{ij'} $ in expression \eqref{DMUnif}.\\

\begin{corollary} \label{cor}
In the particular case where a \textbf{Symmetric Triangular distribution} 
is assumed within the intervals $ I_{ik} $, $ k=j,j' $, for the observation $s_i$ 
of the variables $Y_{j}$ and $Y_{j'}$, the \textbf{square of the Mallows distance} 
between intervals simplifies to:
\begin{equation} \label{DMTSimétrica}
D^{2}_M(\Psi_{Y_{ij}}^{-1}(t),\Psi_{Y_{ij'}}^{-1}(t))=\displaystyle (c_{ij}-c_{ij'})^2+\dfrac{1}{6}(r_{ij}-r_{ij'})^2,
\end{equation}
where $\Psi_{Y_{ij}}^{-1}$ and $\Psi_{Y_{ij'}}^{-1}$ 
are the quantile functions representing $Y_{ij}$ and $Y_{ij'}$ respectively. 
\end{corollary}

\section{Factor Analysis of Interval Data}
\label{sec:3}

\subsection{Factor Model}
\label{sec3:1}

Let $Y_{1},Y_{2},\ldots,Y_{p}$ be the observed interval-valued variables measured on a set of n units $S=\{s_1,\ldots,s_n\}$, as exemplified in Table 1 of Section 2. To specify the factor model we will use the standardized interval-valued variable $Z_{1},Z_{2},\ldots,Z_{p}$ defined by $Z_{ij}= \Big[\frac{l_{ij}-\overline{Y}_{j}}{S_{j}}, \frac{u_{ij}-\overline{Y}_{j}}{S_{j}}\Big]$. All variables $Z_{j}$, $ j = 1,\ldots,p $, have null sample mean and unit sample variance. 

The proposed factor model presumes that these variables are linearly dependent 
on few unobservable interval-valued variables $f_{1},f_{2},\ldots,f_{m}$ $(m<<p)$ 
called \textbf{common factors} and $p$ interval-valued sources of variation 
$\varepsilon_{1},\varepsilon_{2},\ldots,\varepsilon_{p}$ 
called \textbf{specific factors}, or \textbf{errors}, such that
\begin{equation} \label{ModeloFat_1}
{Z_{j}}= \ell_{j1}f_{1}+\ell_{j2}f_{2}+\ldots+\ell_{jm}f_{m}+
\varepsilon_{j}, \quad j = 1,\ldots,p. 
\end{equation}
where $\ell_{jk}$'s, $k=1,\ldots,m$, are the model coefficients, 
real values, usually termed as \textbf{factor loadings} \cite{Johnson2002,Johnson1998}.
The interval $\varepsilon_{j}$ describes the residual variation specific to the $j$th variable $Z_{j}$ and its variance, $S^2_{\varepsilon_{j}}$, is called the \textbf{specific variance} of the $j$th variable.\\
\\
If we replace in the previous model \eqref{ModeloFat_1}, each interval by the associated quantile function 
we obtain the model rewritten as follows,
\begin{equation} \label{ModeloFat_2}
\Psi_{Z_{j}}^{-1}(t)= \underbrace{\ell_{j1}\Psi_{f_{1}}^{-1}(*)+\ell_{j2}\Psi_{f_{2}}^{-1}(*)+\ldots+\ell_{jm}\Psi_{f_{m}}^{-1}(*)}_{\Psi_{CL_{j}}^{-1}}+\Psi_{\varepsilon_{j}}^{-1}(t), \: j = 1,\ldots,p, \: 0 \leq t \leq 1
\end{equation}
with $ * = t $ if $\ell_{ji}>0 $ and $ * = 1-t $ if $\ell_{ji}<0 $, where \vspace{0,15cm} \\ 
\vspace{0,1cm}
\hspace{0,3cm} $\Psi_{Z_{j}}^{-1}$ is the quantile function associated with the standardized interval-valued variable $Z_{j}$; \\ \vspace{0,1cm}
\hspace{0,3cm} $\Psi_{f_{k}}^{-1}$ is the quantile function associated with the interval-valued variable $f_{k}$; \\ \vspace{0,1cm}
\hspace{0,3cm}  $\Psi_{\varepsilon_{j}}^{-1}$ is the quantile function associated with the interval-valued variable $\varepsilon_{j}$.\\
\\ 
\vspace{0,15cm}
In the previous model it is necessary to assume that:\\
\vspace{0,1cm}
\hspace{0,3cm} \textsc{(i)} the $Z_{j}$ variables, the common factors $f_{k}$'s and specific factors $\varepsilon_{j}$'s have null mean;\\
\vspace{0,1cm}
\hspace{0,3cm} \textsc{(ii)} the common factors $f_{k}$'s and specific factors $\varepsilon_{j}$'s are uncorrelated for all combinations of \textit{k} and \textit{j};\\
\vspace{0,1cm}
\hspace{0,3cm} \textsc{(iii)} the common factors $f_{k}$'s are uncorrelated and have unit variance. \\

These assumptions and the preceding model constitute the \textit{orthogonal factor model}. 
If (III) is not verified we have the so-called \textit{model of oblique factors}. 
We can thus say that the orthogonal model of factor analysis assumes that there is a smaller set of uncorrelated interval-valued variables that explain the relations between the observed interval-valued variables.\\
\\
As in the classic case, the factors may be extracted by different methods. Here we consider extraction by Principal Component and by Principal Axis Factoring on the interval-valued variables correlation matrix. 

Principal Component is the perhaps the most commonly used extraction method, but it implicitly assumes that all communalities are equal to one, so that the variables' variances could (theorethically) be completely explained by common factors. Principal Axis Factoring, on the other hand, assumes a model with common and unique factors, and therefore variance cannot be explained just by common factors; the method proceeds iteratively, by first estimating communalities and then trying to identify the common factors responsible for these communalities and the correlations between variables (see, e.g. \cite{Sharma1996}). 
Therefore, a factor model  is implicitly assumed in the Principal Axis Factoring.\\
In both cases, and taking into account the model assumptions,\\
\vspace{0,1cm}
\hspace{0,3cm} \textsc{(i)} the correlation between $Z_{j}$ and $f_{k}$, denoted by $ Corr(Z_{j},f_{k}) $, is $\ell_{jk}$, the loading of the $j$th variable on the $k$th factor. For this reason it is said that each loading $\ell_{jk}$ measures the contribution of the $k$th common factor to the $j$th variable;\\
\vspace{0,1cm}
\hspace{0,3cm} \textsc{(ii)} the variance of $Z_{j}$ can be partitioned as $ S^2_{Z_{j}} = \displaystyle \sum_{k=1}^{m} \ell^2_{jk} + S^2_{\varepsilon_{j}} = 1 $, and the proportion of the variance of $Z_{j}$ that is explained by the common factors, $ \displaystyle \sum_{k=1}^{m} \ell^2_{jk} $, is named the \textbf{communality} of the $j$th variable;\\
\vspace{0,1cm}
\hspace{0,3cm} \textsc{(iii)} the correlation between $Z_{j}$ and $Z_{j'}$ is $ Corr(Z_{j},Z_{j'}) = \displaystyle \sum_{k=1}^{m} \ell_{jk}\ell_{j'k} $.

\subsection{Factor Scores}
\label{sec3:2}

In this section we will present two approaches to interval-valued factor scores estimation, 
inspired in methods for real-valued data, namely, the Bartlett and the Anderson-Rubin methods \cite{DiStefano2009}.

The method suggested by Bartlett, also known as the Weighted Least Squares method, 
chooses as estimates of factor scores those that minimize the sum of squared errors, 
weighted by the reciprocal of their variances. It can be shown \cite{Johnson1998} that for real-valued variables the resulting factor scores are nothing more than the values of the (scaled) principal components.

Our first proposal, inspired by this idea, is to consider the sum of the squared Mallows distances between $ \Psi_{Z_{j}}^{-1} $ and $ \Psi_{CL_{j}}^{-1} $, taking into account model \eqref{ModeloFat_2} and choose the interval-valued factor scores estimates that minimize that sum, weighted by the reciprocal of the interval variable $ \varepsilon_{j} $ variance, $ S^2_{\varepsilon_{j}} $, that is,
\begin{center}
Minimize $ \quad \displaystyle \sum_{j=1}^{p}\dfrac{D^{2}_M(\Psi_{Z_{j}}^{-1},\Psi_{CL_{j}}^{-1})}{S^2_{\varepsilon_{j}}}$.\\
\end{center}

It is important to underline that the factor scores are no longer the values of the (scaled) principal components and, to the best of our knowledge, cannot be obtained by a closed formula.\\

The method proposed by Anderson and Rubin adapts the approach of Bartlett 
such that the factor scores are not only uncorrelated with other factors, 
but also uncorrelated with each other. Thus, our second proposal is to 
\begin{center}
Minimize $ \quad  \displaystyle \sum_{j=1}^{p}\dfrac{D^{2}_M(\Psi_{Z_{j}}^{-1},\Psi_{CL_{j}}^{-1})}{S^2_{\varepsilon_{j}}}$
\end{center}
subject to the condition $Corr(\widehat{f}_{k},\widehat{f}_{k'})=0$, for $ k \ne k' $, $ \forall \; k,k'=1,...,m $.\\

In both approaches the estimates are obtained by solving an optimization problem.

In order to find the factor scores, we relied in the optimization routines of the R system. In particular, in the 'Bartlett method', for each unit we specified an error function, \textit{SumDist}, for the weighted sum of Mallows distances, which takes as its arguments the relevant distribution parameters. Then we minimize \textit{SumDist} by the \textit{nlminb} routine of the R system, using as starting points the U(0,1) (Uniform distribution) and the Tr(0,1,2) (Triangular distribution) for the \textit{nlminb} search.  We have found that convergence was usually obtained whitin a few dozen iterations. To check for potential problems created by local optima, we conducted some experiments with different starting points, and concluded that our procedure was robust, with the search converging always to the same solutions, even after large perturbations of the search origin.
For the 'Anderson-Rubin method', we defined a global error function, \textit{SumDistFactort}, that adds the sum of weighted Mallows distances for all entities with sum of squared correlations between factor scores, multiplied by a large penalty. We used again the \textit{nlminb} routine using as starting points the parameters of distributions found by the 'Bartlett method'. However, in this case we found evidence of local optima, and in order to mitigate this dependence we repeted the local search for different starting points until the best solution found did not change after many different iterations of this procedure.
\\

\section{Synthetic Data}
\label{sec:4}

In this section we analyse the behaviour of the proposed method on synthetic data with predefined correlation structures.
We consider cases where all interval-valued variables are highly or only moderatly correlated and cases where there are differents blocks of highly and$/$or moderatly correlated variables. Is assumed high correlations if values are between 0.8 and 1 and moderate correlations between 0.5 and 0.8.\\
The generation of the synthetic data was done in three main steps:
\begin{enumerate}
	\item  Generate two different matrices with similar correlation structures: the correlation matrix between the centers R$_{c}$ and the correlation matrix between the half-ranges R$_{r}$. These correlation matrices were generated by application of Algorithm 1 suggested by Hardin. For more details on correlation matrices simulation with or without a given structure, see Hardin \cite{Hardin2013}.
	\item Generate the matrix of the centers of the intervals as the product of two matrices: $C_{ini}\times L_{C}$ where,
		\begin{itemize}
			\item R$_{c}$ = $L_{C}^{t}$ $\times$ $L_{C}$ is the Choleski decomposition of R$_{c}$.
			\item elements of matrix $C_{ini}$ are randomly selected from a Uniform distribution in the interval $(a, b)$ such that $a \frown U (0,5)$ and $b \frown U (5,15)$.
		\end{itemize}
	\item Generate the matrix of the half-ranges of the intervals as the product of two matrices: $R_{ini}\times L_{R}$ where,
		\begin{itemize}
			\item R$_{r}$ = $L_{R}^{t}$ $\times$ $L_{R}$ is the Choleski decomposition of R$_{r}$.
			\item elements of matrix $R_{ini}$ are randomly selected from a Uniform distribution in the interval $[0.1, 1]$.
		\end{itemize}
\end{enumerate}

Below, we define 6 different correlation matrix structures and present the correlation matrices generated between the centers R$_{c}$ and between the half-ranges R$_{r}$ for each of the cases. Problems with 10 interval-valued variables are analysed. In each case a set of 100 values of centers and half-ranges are generated.\\

Case 1: All variables highly correlated. 
	\begin{small}
		{\begin{equation*}
			R_{c}=\left[
			\begin{array}{cccccccccc}
				1 & 0.898 & 0.914 & 0.910 & 0.915 & 0.891 & 0.890 & 0.907 & 0.909 & 0.892 \\
				& 1 & 0.893 & 0.920 & 0.907 & 0.871 & 0.907 & 0.915 & 0.902 & 0.920\\
				& & 1 & 0.907 & 0.920 & 0.912 & 0.915 & 0.919 & 0.920 & 0.903 \\
				& & & 1 & 0.912 & 0.899 & 0.905 & 0.872 & 0.888 & 0.931 \\
				& & & & 1 & 0.906 & 0.917 & 0.920 & 0.938 & 0.928 \\
				& & & & & 1 & 0.926 & 0.929 & 0.931 & 0.888 \\
				& & & & & & 1 & 0.912 & 0.925 & 0.922 \\
				& & & & & & & 1 & 0.943 & 0.915 \\
				& & & & & & & & 1 & 0.903 \\
				& & & & & & & & & 1 
			\end{array}
			\right]
		\end{equation*} }
		{\begin{equation*}
				R_{r}=\left[
				\begin{array}{cccccccccc}
				1 & 0.818 & 0.888 & 0.919 & 0.862 & 0.861 & 0.883 & 0.887 & 0.911 & 0.910 \\
				& 1 & 0.852 & 0.825 & 0.837 & 0.779 & 0.856 & 0.782 & 0.830 & 0.797 \\
				& & 1 & 0.910 & 0.913 & 0.824 & 0.929 & 0.901 & 0.921 & 0.884 \\
				& & & 1 & 0.926 & 0.853 & 0.918 & 0.878 & 0.918 & 0.902 \\
				& & & & 1 & 0.878 & 0.996 & 0.902 & 0.910 & 0.864 \\
				& & & & & 1 & 0.887 & 0.894 & 0.881 & 0.895 \\
				& & & & & & 1 & 0.872 & 0.927 & 0.887 \\
				& & & & & & & 1 & 0.918 & 0.897 \\
				& & & & & & & & 1 & 0.912 \\
				& & & & & & & & & 1
				\end{array}
				\right]
		\end{equation*}}
	\end{small}

\vspace{2cm}

Case 2: All variables moderatly correlated.

	\begin{small}
			{\begin{equation*}
				R_{c}=\left[
				\begin{array}{cccccccccc}
				1 & 0.661 & 0.700 & 0.611 & 0.623 & 0.693 & 0.706 & 0.773 & 0.705 & 0.709 \\
				& 1 & 0.768 & 0.686 & 0.683 & 0.695 & 0.685 & 0.726 & 0.719 & 0.723\\
				& & 1 & 0.662 & 0.667 & 0.735 & 0.697 & 0.781 & 0.667 & 0.748 \\
				& & & 1 & 0.726 & 0.686 & 0.760 & 0.686 & 0.677 & 0.660 \\
				& & & & 1 & 0.688 & 0.714 & 0.652 & 0.779 & 0.724 \\
				& & & & & 1 & 0.612 & 0.763 & 0.671 & 0.715 \\
				& & & & & & 1 & 0.649 & 0.703 & 0.707 \\
				& & & & & & & 1 & 0.723 & 0.665 \\
				& & & & & & & & 1 & 0.629 \\
				& & & & & & & & & 1
				\end{array}
				\right]
				\end{equation*} }
			
			{\begin{equation*}
				R_{r}=\left[
				\begin{array}{cccccccccc}
				1 & 0.760 & 0.715 & 0.767 & 0.730 & 0.564 & 0.721 & 0.464 & 0.711 & 0.798 \\
				& 1 & 0.634 & 0.691 & 0.762 & 0.644 & 0.710 & 0.546 & 0.716 & 0.708 \\
				& & 1 & 0.759 & 0.695 & 0.534 & 0.694 & 0.536 & 0.768 & 0.696 \\
				& & & 1 & 0.735 & 0.601 & 0.711 & 0.496 & 0.757 & 0.817 \\
				& & & & 1 & 0.683 & 0.766 & 0.677 & 0.798 & 0.710 \\
				& & & & & 1 & 0.573 & 0.458 & 0.619 & 0.617 \\
				& & & & & & 1 & 0.548 & 0.686 & 0.751 \\
				& & & & & & & 1 & 0.573 & 0.475 \\
				& & & & & & & & 1 & 0.727 \\
				& & & & & & & & & 1
				\end{array}
				\right]
			\end{equation*}}
	\end{small}

Case 3: Two blocks of highly correlated variables. 

	\begin{small}
			{\begin{equation*}
				R_{c}=\left[
				\begin{array}{cccccccccc}
				1 & 0.858 & 0.891 & 0.220 & 0.203 & 0.230 & 0.170 & 0.234 & 0.228 & 0.192 \\
				& 1 & 0.893 & 0.183 & 0.216 & 0.206 & 0.210 & 0.173 & 0.213 & 0.173 \\
				& & 1 & 0.134 & 0.196 & 0.193 & 0.245 & 0.223 & 0.235 & 0.159 \\
				& & & 1 & 0.831 & 0.805 & 0.792 & 0.806 & 0.797 & 0.808 \\
				& & & & 1 & 0.824 & 0.813 & 0.799 & 0.793 & 0.821 \\
				& & & & & 1 & 0.796 & 0.865 & 0.792 & 0.770 \\
				& & & & & & 1 & 0.818 & 0.786 & 0.812 \\
				& & & & & & & 1 & 0.807 & 0.808 \\
				& & & & & & & & 1 & 0.779 \\
				& & & & & & & & & 1
				\end{array}
				\right]
				\end{equation*} }

			{\begin{equation*}
				R_{r}=\left[
				\begin{array}{cccccccccc}
				1 & 0.843 & 0.829 & 0.254 & 0.238 & 0.249 & 0.233 & 0.234 & 0.281 & 0.255 \\
				& 1 & 0.877 & 0.238 & 0.212 & 0.229 & 0.248 & 0.264 & 0.272 & 0.256 \\
				& & 1 & 0.267 & 0.276 & 0.260 & 0.221 & 0.260 & 0.259 & 0.260 \\
				& & & 1 & 0.918 & 0.857 & 0.880 & 0.926 & 0.906 & 0.900 \\
				& & & & 1 & 0.941 & 0.918 & 0.900 & 0.904 & 0.890 \\
				& & & & & 1 & 0.896 & 0.866 & 0.919 & 0.890 \\
				& & & & & & 1 & 0.866 & 0.873 & 0.885 \\
				& & & & & & & 1 & 0.917 & 0.906 \\
				& & & & & & & & 1 & 0.914 \\
				& & & & & & & & & 1
				\end{array}
				\right]
				\end{equation*}}
	\end{small}

Case 4: Two blocks of moderatly correlated variables. 

	\begin{small}
			{\begin{equation*}
				R_{c}=\left[
				\begin{array}{cccccccccc}
				1 & 0.541 & 0.529 & 0.551 & 0.068 & 0.102 & 0.048 & 0.097 & 0.094 & 0.118 \\
				& 1 & 0.567 & 0.570 & 0.095 & 0.093 & 0.095 & 0.127 & 0.117 & 0.129 \\
				& & 1 & 0.573 & 0.142 & 0.059 & 0.099 & 0.103 & 0.145 & 0.095 \\
				& & & 1 & 0.059 & 0.058 & 0.098 & 0.130 & 0.108 & 0.087 \\
				& & & & 1 & 0.624 & 0.674 & 0.625 & 0.673 & 0.636 \\
				& & & & & 1 & 0.645 & 0.649 & 0.642 & 0.634 \\
				& & & & & & 1 & 0.653 & 0.606 & 0.635 \\
				& & & & & & & 1 & 0.594 & 0.650 \\
				& & & & & & & & 1 & 0.640 \\
				& & & & & & & & & 1
				\end{array}
				\right]
				\end{equation*} }

			{\begin{equation*}
				R_{r}=\left[
				\begin{array}{cccccccccc}
				1 & 0.566 & 0.629 & 0.577 & 0.192 & 0.192 & 0.238 & 0.222 & 0.215 & 0.205 \\
				& 1 & 0.580 & 0.594 & 0.205 & 0.224 & 0.185 & 0.232 & 0.194 & 0.217\\
				& & 1 & 0.567 & 0.215 & 0.245 & 0.189 & 0.217 & 0.231 & 0.197 \\
				& & & 1 & 0.199 & 0.224 & 0.158 & 0.202 & 0.215 & 0.169 \\
				& & & & 1 & 0.631 & 0.608 & 0.648 & 0.626 & 0.648 \\
				& & & & & 1 & 0.611 & 0.622 & 0.654 & 0.623 \\
				& & & & & & 1 & 0.611 & 0.619 & 0.640 \\
				& & & & & & & 1 & 0.618 & 0.660 \\
				& & & & & & & & 1 & 0.662 \\
				& & & & & & & & & 1
				\end{array}
				\right]
				\end{equation*}}
	\end{small}

Case 5: One block of highly correlated variables and one block of moderatly correlated variables. 

	\begin{small}
			{\begin{equation*}
				R_{c}=\left[
				\begin{array}{cccccccccc}
				1 & 0.909 & 0.956 & 0.104 & 0.122 & 0.096 & 0.121 & 0.091 & 0.141 & 0.092 \\
				& 1 & 0.913 & 0.082 & 0.130 & 0.133 & 0.108 & 0.080 & 0.076 & 0.088 \\
				& & 1 & 0.111 & 0.126 & 0.097 & 0.082 & 0.0780 & 0.102 & 0.134 \\
				& & & 1 & 0.587 & 0.564 & 0.595 & 0.563 & 0.617 & 0.560 \\
				& & & & 1 & 0.599 & 0.562 & 0.587 & 0.613 & 0.557 \\
				& & & & & 1 & 0.552 & 0.581 & 0.564 & 0.599 \\
				& & & & & & 1 & 0.618 & 0.586 & 0.549 \\
				& & & & & & & 1 & 0.533 & 0.569 \\
				& & & & & & & & 1 & 0.544 \\
				& & & & & & & & & 1
				\end{array}
				\right]
				\end{equation*} }

			{\begin{equation*}
				R_{r}=\left[
				\begin{array}{cccccccccc}
				1 & 0.912 & 0.931 & 0.201 & 0.173 & 0.207 & 0.224 & 0.179 & 0.231 & 0.184 \\
				& 1 & 0.938 & 0.204 & 0.195 & 0.204 & 0.181 & 0.221 & 0.230 & 0.180 \\
				& & 1 & 0.218 & 0.183 & 0.194 & 0.230 & 0.189 & 0.202 & 0.236 \\
				& & & 1 & 0.608 & 0.617 & 0.575 & 0.597 & 0.618 & 0.613 \\
				& & & & 1 & 0.626 & 0.610 & 0.631 & 0.587 & 0.606 \\
				& & & & & 1 & 0.602 & 0.620 & 0.601 & 0.576 \\
				& & & & & & 1 & 0.607 & 0.609 & 0.641 \\
				& & & & & & & 1 & 0.551 & 0.620 \\
				& & & & & & & & 1 & 0.590 \\
				& & & & & & & & & 1
				\end{array}
				\right]
				\end{equation*}}
	\end{small}

Case 6: Three blocks of highly correlated variables. 

	\begin{small}
			{\begin{equation*}
				R_{c}=\left[
				\begin{array}{cccccccccc}
				1 & 0.901 & 0.786 & 0.048 & 0.117 & 0.044 & 0.083 & 0.133 & 0.110 & 0.104 \\
				& 1 & 0.849 & 0.089 & 0.116 & 0.086 & 0.130 & 0.149 & 0.107 & 0.117 \\
				& & 1 & 0.120 & 0.080 & 0.128 & 0.126 & 0.143 & 0.077 & 0.121 \\
				& & & 1 & 0.889 & 0.900 & 0.910 & 0.101 & 0.085 & 0.110 \\
				& & & & 1 & 0.900 & 0.940 & 0.131 & 0.177 & 0.114 \\
				& & & & & 1 & 0.867 & 0.067 & 0.116 & 0.066 \\
				& & & & & & 1 & 0.166 & 0.076 & 0.109 \\
				& & & & & & & 1 & 0.881 & 0.948 \\
				& & & & & & & & 1 & 0.929 \\
				& & & & & & & & & 1
				\end{array}
				\right]
				\end{equation*} }

			{\begin{equation*}
				R_{r}=\left[
				\begin{array}{cccccccccc}
				1 & 0.941 & 0.938 & 0.191 & 0.126 & 0.161 & 0.142 & 0.185 & 0.192 & 0.159 \\
				& 1 & 0.945 & 0.174 & 0.135 & 0.163 & 0.130& 0.168 & 0.137 & 0.166 \\
				& & 1 & 0.179 & 0.155 & 0.145 & 0.143 & 0.138 & 0.159 & 0.117 \\
				& & & 1 & 0.846 & 0.867 & 0.828 & 0.175 & 0.135 & 0.134 \\
				& & & & 1 & 0.845 & 0.856 & 0.177 & 0.158 & 0.108 \\
				& & & & & 1 & 0.817 & 0.168 & 0.114 & 0.125 \\
				& & & & & & 1 & 0.183 & 0.208 & 0.177 \\
				& & & & & & & 1 & 0.799 & 0.866 \\
				& & & & & & & & 1 & 0.785 \\
				& & & & & & & & & 1
				\end{array}
				\right] 
				\end{equation*}}
	\end{small}

For each case, a factor analysis according to the proposed model (\ref{ModeloFat_1}) was performed assuming three distinct distributions within each interval: Uniform, Triangular Symmetric and Triangular (with the mode randomly chosen within interval).\\ 
The number of common factors retained was defined according to the rule of eigenvalues greater than one and in line with the cumulative proportion of total variation. Both Principal Component and Principal Axis Factoring leads to the extraction of same number of factors in each case.\\
Table \ref{TabfinalSim} presents the number of interval-valued factors extracted in each case, which was the same for all the assumed distributions. 

\begin{center}
	\begin{table} [h!] 
	\caption{ - Number of factors extracted considering the Uniform, Triangular Symmetric and Triangular distribution.}
	\begin{tabular}{|c||c|c|c|c|}
		\hline 
		& Number &   &  &  \\
		& of factors &  1st factor & 2nd factor & 3rd factor  \\  
		\hline \hline
		Case 1 & 1 & All & ---- & ----   \\
		\hline 
		Case 2 & 1 & All & ---- &----   \\ 
		\hline 
		Case 3 & 2 & Variables of 1st group & Variables of 2nd group & ----   \\
		\hline 
		Case 4 & 2 & Variables of 1st group & Variables of 2nd group & ----   \\ 
		\hline 
		Case 5 & 2 & Variables of 1st group & Variables of 1st and 2nd groups & ----    \\
		\hline 
		Case 6 & 3 & Variables of 1st group & Variables of 2nd group & Variables of 3rd group  \\
		\hline
	\end{tabular} \label{TabfinalSim} 
	\end{table}
\end{center}

We can see from Table 2 that the factor analysis of this data succeeds in recovering their original structure in cases 1, 2, 3, 4 and 6. Case 5, where there are both groups of higly and moderatly correlated variables, is somehow more difficult, and the variables of the strongly correlated group sometimes also appear in the definition of the second factor. Nevertheless, this is not much different from similar data conditions in the factor analysis of classic data, and the basic group correlation structure is still recognized by the analysis.

\section{Application}
\label{sec:5}

In this section, we illustrate the methodology proposed above on a car data set and on meteorological data, for the different alternatives concerning \textbf{(a)} the distribution within the intervals: Uniform and Triangular distributions, \textbf{(b)} the technique of factor extraction: Principal Component and Principal Axis Factoring, \textbf{(c)} and the estimation of factor scores: the Bartlett and the Anderson-Rubin methods.
The number of common factors retained was defined according to the rule of eigenvalues $ \hat{\lambda}_{j} $, $ j = 1,\ldots,p $, greater than one and in line with the cumulative proportion of total variation, $ \dfrac{\sum_{k=1}^{j}\hat{\lambda}_{k}}{p} $.

\subsection{Cars Data}
\label{sec5:1}

A factor analysis was performed on a set of 33 car models described by 8 interval-valued variables: Price, Engine Capacity, Top Speed, Acceleration, Wheelbase, Lenght, Width and Height (see Table \ref{Cars}). 
 
\begin{center}
\begin{table} [h!] 
\caption{ - Cars data set (partial view).}
\begin{tabular}{|c||c|c|c|c|} 
\hline 
 & Price & Engine Capacity & \ldots & Height \\ 
\hline \hline
Alfa 145 & $\left[27806,33596\right]$ & $\left[1370,1910\right]$ & \ldots & $\left[143,143\right]$ \\ 
\hline 
Alfa 156 & $\left[41593,62291\right]$ & $\left[1598,2492\right]$ & \ldots & $\left[142,142\right]$ \\
\hline 
Aston Martin & $\left[260500,460000\right]$ & $\left[5935,5935\right]$ & \ldots & $\left[124,132\right]$ \\   
\hline 
 &  &  &  &  \\
\vdots & \vdots & \vdots & \vdots & \vdots \\
 &  &  &  &  \\  
\hline 
Porsche & $\left[147704,246412\right]$ & $\left[3387,3600\right]$ & \ldots & $\left[130,131\right]$ \\  
\hline 
Rover 25 & $\left[21492,33042\right]$ & $\left[1119,1994\right]$ & \ldots & $\left[142,142\right]$ \\  
\hline 
Passat & $\left[39676,63455\right]$ & $\left[1595,2496\right]$ & \ldots & $\left[146,146\right]$ \\  
\hline 
\end{tabular} \label{Cars}
\end{table}
\end{center}

\subsubsection{Uniform Distribution}
\label{sec5:1:1}

In this section we assume that the values within the observed intervals are distributed according to an Uniform distribution. \\
The following is the sample correlation matrix \textbf{R} obtained from the third definition of covariance $ Cov_3 $, using formula (\ref{Cov3U_1_2}):\\

\begin{footnotesize}
	\begin{scriptsize}
		\hspace{2.5cm} Price EngCap \hspace{-0,1cm} TopSpeed Acceler Wheelbase Lenght \hspace{0,1cm} Width \hspace{0,2cm} Height
	\end{scriptsize}
\begin{equation*}
\mathbf{R}=\left[
\begin{array}{cccccccc}
1 & +0.9580 & +0.8712 & -0.7559 & +0.3732 & +0.5159 & +0.8261 & -0.6776 \\
 & 1 & +0.8659 & -0.7296 & +0.4834 & +0.6260 & +0.8502 & -0.6269 \\
 & & 1 & -0.8768 & +0.3396 & +0.5747 &  +0.8529 & -0.7281 \\
 & & & 1 & -0.3973 & -0.5991 & -0.8138 & +0.6037 \\
 & & & & 1 & +0.8657 & +0.5944 & +0.1581 \\
 & & & & & 1 & +0.7635 & -0.0373 \\
 & & & & & & 1 & -0.5431 \\
 & & & & & & & 1
\end{array}
\right]
\end{equation*}
\end{footnotesize}

Both Principal Component and Principal Axis Factoring of this matrix lead to the extraction of two factors, which together represent 89.9 \% and 86.7 \% of the total variance, respectively.
Table \ref{ResumeUnif} summarizes the estimated factor loadings for each variable in the two interval-valued factors, its eigenvalues, the communality of each variable and the cumulative proportion of total sample variance explained, for both methods.

\begin{table} [h!]
\caption{ - Summary of factor analysis, assuming the Uniform distribution within intervals.}

\begin{tabular}{|l||cc|c||cc|c|}
\hline & \multicolumn{3}{|c||}{\textbf{PCF} } & \multicolumn{3}{|c|}{\textbf{PAF} } \\ 
\hline
 & \multicolumn{2}{|c|}{\textbf{Estimated factor} } &  & \multicolumn{2}{|c|}{\textbf{Estimated factor} } & \\ 
 \textbf{Variable}& \multicolumn{2}{|c|}{\textbf{loadings} } & \textbf{Communalities} & \multicolumn{2}{|c|}{\textbf{loadings} } & \textbf{Communalities} \\ 
 & $f_1$ & $f_2$ & & $f_1$ & $f_2$ &\\ 
\hline
Price & $ -0.9219 $ & $ -0.2059 $ & 0.8923 & $ -0.9113 $ & $ -0.2029 $ & 0.8717\\  
				
Eng Capacity & $ -0.9388 $ & $ -0.0724 $ & 0.8865 & $ -0.9273 $ & $ -0.0748 $ & 0.8655\\  
 
Top Speed & $ -0.9384 $ & $-0.2149 $ & 0.9268 & $ -0.9363 $ & $ -0.2174 $ & 0.9239 \\

Acceleration & +0.8832 & +0.0960 & 0.7893 & +0.8509 & +0.0877 & 0.7317 \\  

Wheelbase & $ -0.5636 $ & +0.7825 & 0.9300 & $ -0.5573 $ & +0.7410 & 0.8596\\
 
Lenght & $ -0.7386 $ & +0.6271 & 0.9387 & $ -0.7391 $ & +0.6203 & 0.9310\\  

Width & $ -0.9483 $ & +0.0998 & 0.9092 & $ -0.9418 $ & +0.0925 & 0.8956\\  

Height & +0.6364 & +0.7145 & 0.9155 & +0.6263 & +0.6803 & 0.8550\\  
\hline
Eingenvalues & 5.5594 & 1.6290 &  & 5.4271 & 1.5069 & \\  
\hline
Cumulative proportion  &  &  &  &  &  & \\
of total sample & 0.6949 & 0.8986 &  & 0.6784 & 0.8668 &\\
variance explained &  &  &  &  &  & \\    
\hline
\end{tabular} \label{ResumeUnif}
\end{table}

Analyzing the values presented in Table \ref{ResumeUnif} we can see that there are very little differences between the results obtained by Principal Component and Principal Axis Factoring. From the figures in Table \ref{ResumeUnif}, we may now write the factor model, which in the case of Principal Axis Factoring is as follows:\\
\vspace{0,1cm}
\hspace{4,6cm}$Price= -0.9113f_{1}-0.2029f_{2}+\varepsilon_{Price}$ \\
\vspace{0,1cm}
\hspace{3,4cm}$Eng Capacity= -0.9273f_{1}-0.0748f_{2}+\varepsilon_{Eng Capacity}$\\
\vspace{0,1cm}
\hspace{3,95cm}$Top Speed= -0.9363f_{1}-0.2174f_{2}+\varepsilon_{Top Speed}$ \\
\vspace{0,1cm}
\hspace{3,5cm}$Acceleration= +0.8509f_{1}+0.0877f_{2}+\varepsilon_{Acceleration}$  \\
\vspace{0,1cm}
\hspace{3,8cm}$Wheelbase= -0.5573f_{1}+0.7410f_{2}+\varepsilon_{Wheelbase}$ \\
\vspace{0,1cm}
\hspace{4,4cm}$Lenght= -0.7391f_{1}+0.6203f_{2}+\varepsilon_{Lenght}$ \\
\vspace{0,1cm}
\hspace{4,5cm}$Width= -0.9418f_{1}+0.0925f_{2}+\varepsilon_{Width}$  \\
\vspace{0,1cm}
\hspace{4,4cm}$Height= +0.6263f_{1}+0.6803f_{2}+\varepsilon_{Height}$
\begin{flushleft}
or, if we represent each interval-valued variable by the respective quantile function, 
\end{flushleft}
\hspace{2,85cm}$\Psi_{Price}^{-1}(t)= -0.9113\Psi_{f_{1}}^{-1}(1-t)-0.2029\Psi_{f_{2}}^{-1}(1-t)+\Psi_{\varepsilon_{Price}}^{-1}(t)$ \\
\vspace{0,1cm}
\hspace{1,75cm}$\Psi_{Eng Capacity}^{-1}(t)= -0.9273\Psi_{f_{1}}^{-1}(1-t)-0.0748\Psi_{f_{2}}^{-1}(1-t)+\Psi_{\varepsilon_{Eng Capacity}}^{-1}(t)$ \\
\vspace{0,1cm}
\hspace{2,2cm}$\Psi_{Top Speed}^{-1}(t)= -0.9363\Psi_{f_{1}}^{-1}(1-t)-0.2174\Psi_{f_{2}}^{-1}(1-t)+\Psi_{\varepsilon_{Top Speed}}^{-1}(t)$ \\
\vspace{0,1cm}
\hspace{1,8cm}$\Psi_{Acceleration}^{-1}(t)= +0.8509\Psi_{f_{1}}^{-1}(t)+0.0877\Psi_{f_{2}}^{-1}(t)+\Psi_{\varepsilon_{Acceleration}}^{-1}(t)$ \\
\vspace{0,1cm}
\hspace{2,05cm}$\Psi_{Wheelbase}^{-1}(t)= -0.5573\Psi_{f_{1}}^{-1}(1-t)+0.7410\Psi_{f_{2}}^{-1}(t)+\Psi_{\varepsilon_{Wheelbase}}^{-1}(t)$ \\
\vspace{0,1cm}
\hspace{2,5cm}$\Psi_{Lenght}^{-1}(t)= -0.7391\Psi_{f_{1}}^{-1}(1-t)+0.6203\Psi_{f_{2}}^{-1}(t)+\Psi_{\varepsilon_{Lenght}}^{-1}(t)$ \\
\vspace{0,1cm}
\hspace{2,6cm}$\Psi_{Width}^{-1}(t)= -0.9418\Psi_{f_{1}}^{-1}(1-t)+0.0925\Psi_{f_{2}}^{-1}(t)+\Psi_{\varepsilon_{Width}}^{-1}(t)$ \\
\vspace{0,1cm}
\hspace{2,5cm}$\Psi_{Height}^{-1}(t)= +0.6263\Psi_{f_{1}}^{-1}(t)+0.6803\Psi_{f_{2}}^{-1}(t)+\Psi_{\varepsilon_{Height}}^{-1}(t)$ \\
with $0 \leq t \leq 1 $. \\

The 1st factor presents high factor loadings for Price, Engine Capacity, Top Speed, Acceleration and Width and explains 67.8\% of the total variance. The 2nd factor, with high factor loading for Wheelbase, explains 18.8\% of total variance. It is noted that the Length and Height have high factor loadings on both factors, reflecting the fact that these characteristics do not contribute to the distinction of car models. Additionally, all communalities are high indicating that the two retained factors are suitable for describing the latent relational structure between characteristics of the car models. \\
We can thus say that the factor model distinguishes the car models with higher price, higher engine capacity, higher top speed, greater width and shorter acceleration time from those with opposite characteristics; furthermore it separates car models with larger wheelbase from the others.\\

The model based on Principal Component extraction may be written in a similar way, using the correspondent values of Table \ref{ResumeUnif}.\\

\begin{figure} [h!]
	\subfigure [`Bartlett method'] {\includegraphics[width=7.5cm, height=7.5cm]{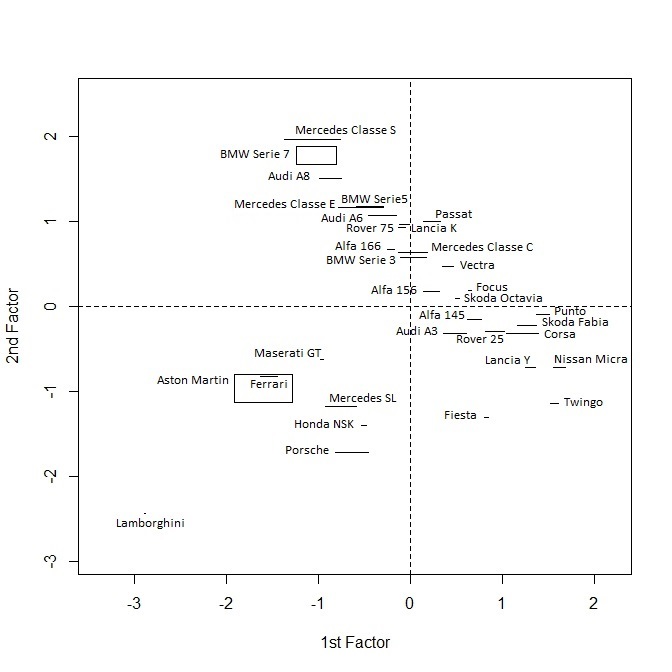}}
	\subfigure [`Anderson-Rubin method'] {\includegraphics[width=7.5cm, height=7.5cm]{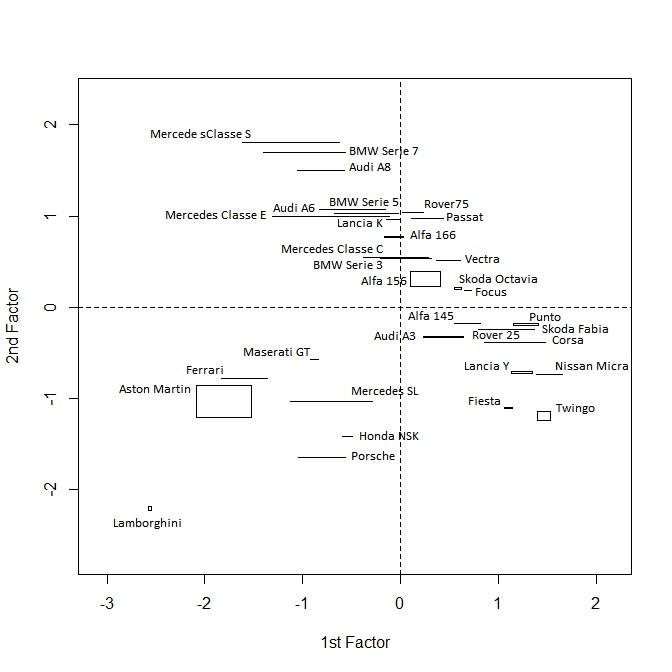}}	
	\caption{ - Factor scores obtained for the car dataset through the model based on Principal Components and considering the Uniform distribution within intervals.} \label{fig_Unif_PCF}
\end{figure}

\begin{figure} [h!] 
	\subfigure [`Bartlett method'] {\includegraphics[width=7.5cm, height=7.5cm]{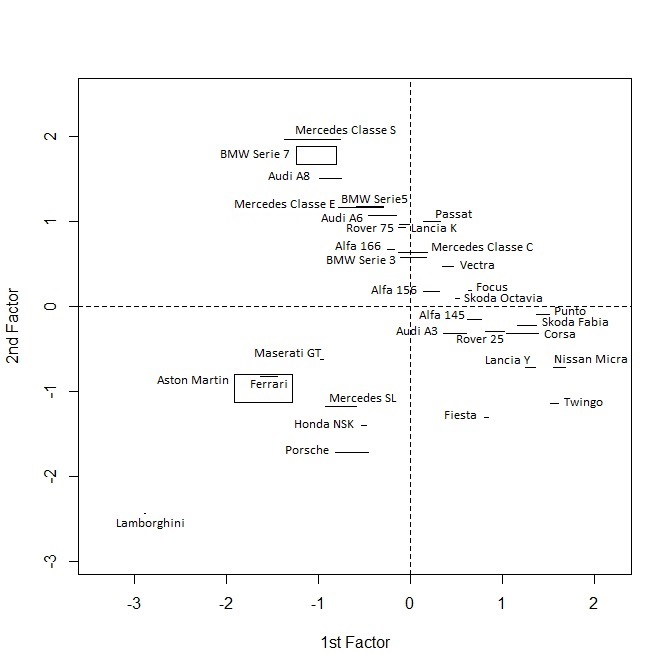}}
	\subfigure [`Anderson-Rubin method'] {\includegraphics[width=7.5cm, height=7.5cm]{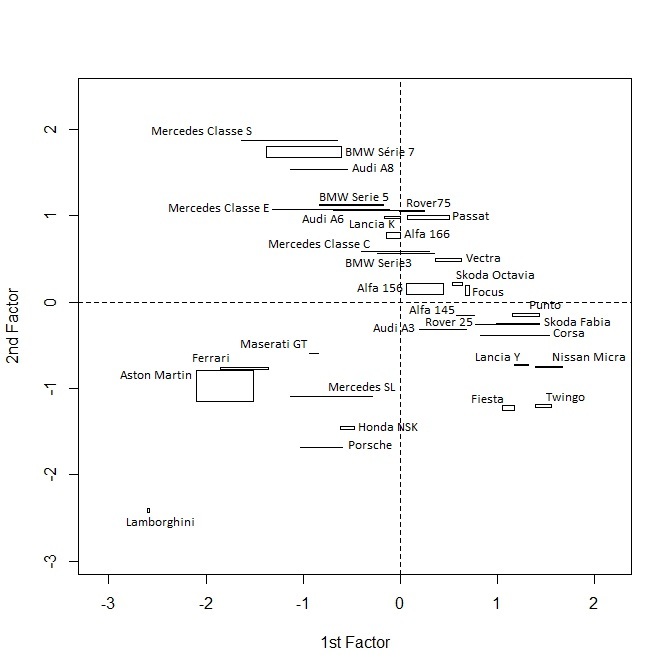}}	
	\caption{ - Factor scores obtained for the car dataset through the model based on Principal Axis Factoring and considering the Uniform distribution within intervals.} \label{fig_Unif_PAF}
\end{figure}

Figures \ref{fig_Unif_PCF} and \ref{fig_Unif_PAF} show the 33 car models in the plane defined by the two interval-valued factors, obtained by the two extracting methods and by the two factor scores estimation methods considerated.
As it can be seen, whereas the first factor distinguishes the upscale car models from the low cost car models, the second factor differentiates essentially car models with greater wheelbase from the smaller ones.

From the observation of the Figures \ref{fig_Unif_PCF} and \ref{fig_Unif_PAF} we can conclude that factor scores obtained by the two methods are very similar. 
However, we note that less degenerate intervals were obtained in the 2nd factor on the factor scores obtained by the model based on Principal Axis Factoring, and this difference is more noticeable when the `Anderson-Rubin method' is choosen.

\subsubsection{Triangular Distribution}
\label{sec5:1:2}

We now assume a Triangular distribution within each observed interval with a randomly generated mode. 
The data may hence be represented by triplets (min, mode, max) as in Table \ref{CarsT}.

\begin{center}
\begin{table} [h!] 
\caption{ - Car data set (partial view) described by triplets (min, mode, max).}
\begin{tabular}{|c||c|c|c|c|} 
\hline 
 & Price & Engine Capacity & \ldots & Height \\ 
\hline \hline
Alfa 145 & $(27806,32566,33596)$ & $(1370,1609,1910)$ & \ldots & $(143,143,143)$ \\ 
\hline 
Alfa 156 & $(41593,61491,62291)$ & $(1598,2249,2492)$ & \ldots & $(142,142,142)$ \\  
\hline
Aston Martin & $(260500,386054,460000)$ & $(5935,5935,5935)$ & \ldots & $(124,131,132)$ \\  
\hline  
 &  &  &  &  \\
\vdots & \vdots & \vdots & \vdots & \vdots \\
 &  &  &  &  \\  
\hline 
Porsche & $(147704,242211,246412)$ & $(3387,3578,3600)$ & \ldots & $(130,130,131)$ \\  
\hline 
Rover 25 & $(21492,29242,33042)$ & $(1119,1532,1994)$ & \ldots & $(142,142,142)$ \\  
\hline 
Passat & $(39676,45063,63455)$ & $(1595,2360,2496)$ & \ldots & $(146,146,146)$ \\  
\hline 
\end{tabular} \label{CarsT}
\end{table}
\end{center}

Applying the covariance definition $ Cov_3 $ as in formula (\ref{Cov3T_2}) 
we obtain the following correlation matrix: \\

\begin{footnotesize}
	\begin{scriptsize}
	\hspace{2.5cm} Price EngCap \hspace{-0,1cm} TopSpeed Acceler Wheelbase Lenght \hspace{0,1cm} Width \hspace{0,2cm} Height
	\end{scriptsize}
\begin{equation*}
\mathbf{R}=\left[
\begin{array}{cccccccc}
1 & +0.9527 & +0.8942 & -0.7920 & +0.3475 & +0.5091 & +0.8415 & -0.7280 \\
 & 1 & +0.8694 & -0.7659 & +0.4897 & +0.6352 & +0.8672 & -0.6275 \\
 & & 1 & -0.9108 & +0.3354 & +0.5720 &  +0.8596 & -0.7326 \\
 & & & 1 & -0.4101 & -0.6035 & -0.8238 & +0.6099 \\
 & & & & 1 & +0.8672 & +0.5912 & +0.1570 \\
 & & & & & 1 & +0.7627 & -0.0346 \\
 & & & & & & 1 & -0.5470 \\
 & & & & & & & 1
\end{array}
\right]
\end{equation*}
\end{footnotesize}

Principal Component and Principal Axis Factoring of this matrix leads to the extraction of two factors, with values of estimated factor loadings in interval-valued factors, eigenvalues and communalities which are very similar. For this reason in the Table \ref{ResumeTriang} we only indicate those values for the Principal Axis Factoring method.

\begin{center}
\begin{table} [h!]
\caption{ - Summary of factor analysis, obtained by Principal Axis Factoring, considering the Triangular distribution within intervals.}
\begin{tabular}{|l||cc|c|}
\hline 
\textbf{Variable} & \multicolumn{2}{|c|}{\textbf{Estimated factor loadings} } & \textbf{Communalities} \\ 
 & $f_1$ & $f_2$ & \\ 
\hline
Price & $ -0.9222 $ & $ -0.2266 $ & 0.9018 \\  

Eng Capacity & $ -0.9327 $ & $ -0.0584 $ & 0.8733 \\  
 
Top Speed & $ -0.9438 $ & $-0.2155 $ & 0.9372 \\

Acceleration & +0.8737 & +0.0836 & 0.7704  \\  

Wheelbase & $ -0.5485 $ & +0.7465 & 0.8581 \\
 
Lenght & $ -0.7327 $ & +0.6312 & 0.9354 \\  

Width & $ -0.9442 $ & +0.0967 & 0.9009 \\  

Height & +0.6310 & +0.6710 & 0.8484 \\  
\hline
Eingenvalues & 5.5019 & 1.5235 &  \\  
\hline
Cumulative proportion of  &  &  &  \\
total sample variance explained & 0.6877 & 0.8782 &  \\    
\hline
\end{tabular} \label{ResumeTriang}
\end{table}
\end{center}

Based on the factor loadings of the model we can conclude that the variables Price, Engine Capacity, Top Speed, Acceleration and Width are strongly related to the 1st factor and weakly associated with the 2nd factor, whereas the variable Wheelbase is strongly associated with the 2nd factor and more weakly associated with the 1st factor. Length and Height have high factor loadings on both factors, so do not contribute to the distinction of car models. 68.8\% of total variance is explained by the 1st factor and 19.0\% by the 2nd factor, which together represent 87.8\% of the total variance. Furthermore, all communalities are high indicating that the two factors retained are suitable for describing the latent relational structure between characteristics of the car models. \\

The resulting factor model is,\\
\vspace{0,1cm}
\hspace{4,6cm}$Price= -0.9222f_{1}-0.2266f_{2}+\varepsilon_{Price}$ \\
\vspace{0,1cm}
\hspace{3,4cm}$Eng Capacity= -0.9327f_{1}-0.0584f_{2}+\varepsilon_{Eng Capacity}$ \\
\vspace{0,1cm}
\hspace{3,95cm}$Top Speed= -0.9438f_{1}-0.2155f_{2}+\varepsilon_{Top Speed}$ \\
\vspace{0,1cm}
\hspace{3,5cm}$Acceleration= +0.8737f_{1}+0.0836f_{2}+\varepsilon_{Acceleration}$ \\
\vspace{0,1cm}
\hspace{3,8cm}$Wheelbase= -0.5485f_{1}+0.7465f_{2}+\varepsilon_{Wheelbase}$ \\
\vspace{0,1cm}
\hspace{4,4cm}$Lenght= -0.7327f_{1}+0.6312f_{2}+\varepsilon_{Lenght}$ \\
\vspace{0,1cm}
\hspace{4,5cm}$Width= -0.9442f_{1}+0.0967f_{2}+\varepsilon_{Width}$  \\
\vspace{0,1cm}
\hspace{4,4cm}$Height= +0.6410f_{1}+0.6710f_{2}+\varepsilon_{Height}$
\begin{flushleft}
or, if we represent each interval-valued variable by the respective quantile function, \vspace{0,1cm}
\end{flushleft}
\vspace{0,1cm}
\hspace{2,85cm}$\Psi_{Price}^{-1}(t)= -0.9222\Psi_{f_{1}}^{-1}(1-t)-0.2266\Psi_{f_{2}}^{-1}(1-t)+\Psi_{\varepsilon_{Price}}^{-1}(t)$ \\
\vspace{0,1cm}
\hspace{1,75cm}$\Psi_{Eng Capacity}^{-1}(t)= -0.9327\Psi_{f_{1}}^{-1}(1-t)-0.0584\Psi_{f_{2}}^{-1}(1-t)+\Psi_{\varepsilon_{Eng Capacity}}^{-1}(t)$ \\
\vspace{0,1cm}
\hspace{2,2cm}$\Psi_{Top Speed}^{-1}(t)= -0.9438\Psi_{f_{1}}^{-1}(1-t)-0.2155\Psi_{f_{2}}^{-1}(1-t)+\Psi_{\varepsilon_{Top Speed}}^{-1}(t)$ \\
\vspace{0,1cm}
\hspace{1,8cm}$\Psi_{Acceleration}^{-1}(t)= +0.8737\Psi_{f_{1}}^{-1}(t)+0.0836\Psi_{f_{2}}^{-1}(t)+\Psi_{\varepsilon_{Acceleration}}^{-1}(t)$ \\
\vspace{0,1cm}
\hspace{2,1cm}$\Psi_{Wheelbase}^{-1}(t)= -0.5485\Psi_{f_{1}}^{-1}(1-t)+0.7465\Psi_{f_{2}}^{-1}(t)+\Psi_{\varepsilon_{Wheelbase}}^{-1}(t)$ \\
\vspace{0,1cm}
\hspace{2,5cm}$\Psi_{Lenght}^{-1}(t)= -0.7327\Psi_{f_{1}}^{-1}(1-t)+0.6312\Psi_{f_{2}}^{-1}(t)+\Psi_{\varepsilon_{Lenght}}^{-1}(t)$ \\
\vspace{0,1cm}
\hspace{2,6cm}$\Psi_{Width}^{-1}(t)= -0.9442\Psi_{f_{1}}^{-1}(1-t)+0.0967\Psi_{f_{2}}^{-1}(t)+\Psi_{\varepsilon_{Width}}^{-1}(t)$ \\
\vspace{0,1cm}
\hspace{2,5cm}$\Psi_{Height}^{-1}(t)= +0.6310\Psi_{f_{1}}^{-1}(t)+0.6710\Psi_{f_{2}}^{-1}(t)+\Psi_{\varepsilon_{Height}}^{-1}(t)$ 
\begin{flushleft}
with $0 \leq t \leq 1 $.\\
\end{flushleft}
\vspace{0,2cm}

\begin{figure} [h!] 
	\subfigure[ref1] [`Bartlett method'] {\includegraphics[width=7.5cm, height=7.5cm]{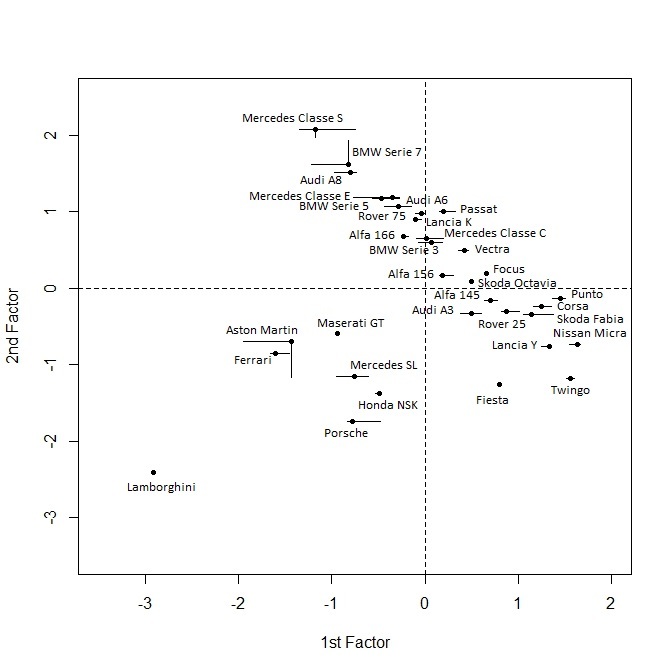}}
	\subfigure[ref2] [`Anderson-Rubin method'] {\includegraphics[width=7.5cm, height=7.5cm]{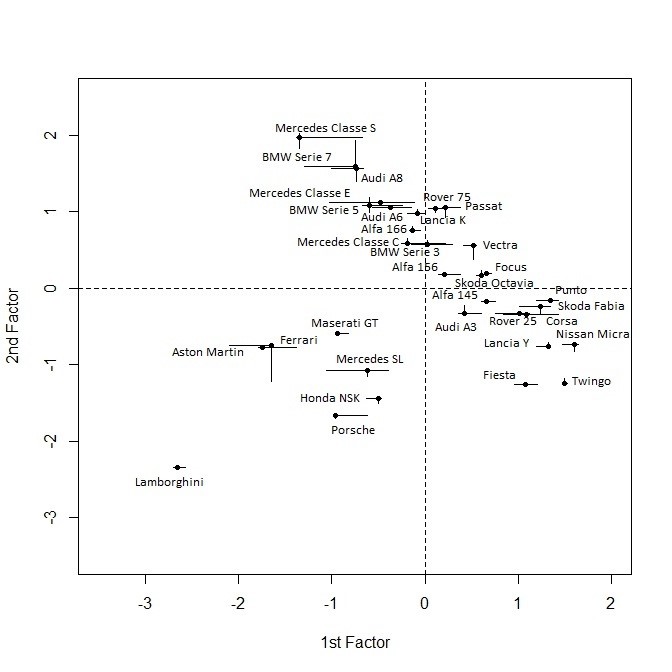}}	
	\caption{ - Factor scores obtained for the car dataset through the model based on Principal Axis Factoring and considering the Triangular distribution within intervals.} \label{fig_Tri_PAF}
\end{figure}

The factor scores obtained by the `Bartlett method' and by the `Anderson-Rubin method' are represented in Figure \ref{fig_Tri_PAF} showing the 33 car models in the plane defined by the two interval-valued factors. From their observation we can conclude that factor scores obtained by the two methods are very similar in the 1st fator, while much less degenerate intervals were obtained in the 2nd factor by `Anderson-Rubin method'.

As it can be observed, the position in the plan of car models assuming the Triangular distribution is very similar to that obtained when the Uniform distribution was assumed, and therefore the conclusions are analogous.
Essentially, the 1st factor distinguishes upscale from low cost car models and the 2nd factor differentiates car models with greater wheelbase from smaller ones.\\

\subsection{Meteorological Data}
\label{sec5:2}
 
In this section a factor analysis is performed on a set of 283 cities of the United States of America described by 13 interval-valued variables: the temperatures (in Fahrenheit degrees) of the 12 months of the year and the annual precipitation (in mm) between the years 1971 e 2000 (see Table \ref{TempPrcUSA}). 

\begin{center}
	\begin{table} [h!] 
		\caption{ - Meteorological data set (partial view).}
		\begin{tabular}{|c||c|c|c|c|c|} 
			\hline 
			& January & February & \ldots & December & Precipitation \\ 
			\hline \hline
			BIRMINGHAM AP, AL & $\left[32.3,52.8\right]$ & $\left[35.4,58.3\right]$ & \ldots & $\left[35.2,56.0\right]$ & $\left[41.8,90.6\right]$ \\ 
			\hline 
			HUNTSVILLE, AL & $\left[30.7,48.9\right]$ & $\left[34.0,54.6\right]$ & \ldots & $\left[33.8,52.4\right]$ & $\left[40.7,89.4\right]$ \\
			\hline 
			MOBILE, AL & $\left[39.5,60.7\right]$ & $\left[42.4,64.5\right]$ & \ldots & $\left[41.6,62.9\right]$ & $\left[47.8,91.2\right]$ \\
			\hline 
			MONTGOMERY, AL & $\left[35.5,57.6\right]$ & $\left[38.6,62.4\right]$ & \ldots & $\left[37.6,60.3\right]$ & $\left[43.5,92.7\right]$ \\
			\hline 
			ANCHORAGE, AK & $\left[9.3,22.2\right]$ & $\left[11.7,25.8\right]$ & \ldots & $\left[11.4,23.7\right]$ & $\left[15.9,65.3\right]$ \\   
			\hline 
			&  &  &  &  & \\
			\vdots & \vdots & \vdots & \vdots & \vdots & \vdots \\
			&  &  &  &  & \\  
			\hline 
			WAKE ISLAND, PC & $\left[73.1,82.4\right]$ & $\left[72.4,82.1\right]$ & \ldots & $\left[74.7,83.9\right]$ & $\left[76.3,88.8\right]$ \\  
			\hline 
			YAP, W CAROLINE IS., PC & $\left[73.7,86.5\right]$ & $\left[73.8,86.7\right]$ & \ldots & $\left[74.2,87.0\right]$ & $\left[73.7,87.7\right]$  \\  
			\hline 
			SAN JUAN, PR & $\left[70.8,82.4\right]$ & $\left[70.9,82.8\right]$ & \ldots & $\left[72.1,83.2\right]$ & $\left[74.0,87.8\right]$ \\  
			\hline 
		\end{tabular} \label{TempPrcUSA}
	\end{table}
\end{center}

\subsubsection{Uniform Distribution}
\label{sec5:2:1}

In this section it is assumed that the values within the observed intervals are distributed according to an Uniform distribution. \\
The following is the sample correlation matrix \textbf{R} obtained from the third definition of covariance $ Cov_3 $, using formula (\ref{Cov3U_1_2}):\\

\begin{footnotesize}
	\hspace{1.1cm} Jan 	\hspace{0.1cm} Feb \hspace{0.13cm} Mar \hspace{0.2cm} Apr \hspace{0.2cm} May \hspace{0.2cm} Jun \hspace{0.25cm} Jul \hspace{0.25cm} Aug \hspace{0.2cm} Sept \hspace{0.25cm} Oct \hspace{0.25cm} Nov \hspace{0.2cm} Dec \hspace{0.2cm} Prec
	\begin{equation*}
	\mathbf{R}=\left[
	\begin{array}{ccccccccccccc}
	1 & 0.9942 & 0.9701 & 0.9136 & 0.8232 & 0.7100 & 0.6437 & 0.7029 & 0.8406 & 0.9274 & 0.9745 & 0.9954 & 0.7310 \\
	& 1 & 0.9875 & 0.9414 & 0.8588 & 0.7549 & 0.6943 & 0.7501 & 0.8721 & 0.9470 & 0.9760 & 0.9886 & 0.7620\\
	& & 1 & 0.9808 & 0.9217 & 0.8372 & 0.7827 & 0.8294 & 0.9254 & 0.9756 & 0.9794 & 0.9725 & 0.8150 \\
	& & & 1 & 0.9772 & 0.9198 & 0.8731 & 0.9054 & 0.9683 & 0.9855 & 0.9556 & 0.9262 & 0.8720 \\
	& & & & 1 & 0.9785 & 0.9438 & 0.9588 & 0.9841 & 0.9595 & 0.8945 & 0.8441 & 0.9003 \\
	& & & & & 1 & 0.9873 & 0.9861 & 0.9688 & 0.9023 & 0.8034 & 0.7346 & 0.8950 \\
	& & & & & & 1 & 0.9944 & 0.9483 & 0.8604 & 0.7436 & 0.6665 & 0.8911 \\
	& & & & & & & 1 & 0.9709 & 0.8996 & 0.7949 & 0.7236 & 0.9052 \\
	& & & & & & & & 1 & 0.9733 & 0.9091 & 0.8584 & 0.9092 \\
	& & & & & & & & & 1 & 0.9757 & 0.9427 & 0.8749 \\
	& & & & & & & & & & 1 & 0.9885 & 0.7958 \\
	& & & & & & & & & & & 1 & 0.7462 \\
	& & & & & & & & & & & & 1 \\
	\end{array}
	\right]
	\end{equation*}
\end{footnotesize}

Both Principal Component and Principal Axis Factoring of this matrix leads to the extraction of two factors, with values of estimated factor loadings in interval-valued factors, eigenvalues and communalities nearly equal. For this reason  we only indicate those values for the Principal Axis Factoring method in Table \ref{ResumeUnifMeteo}. 

\begin{center}
	\begin{table} [h!]
		\caption{ - Summary of factor analysis, obtained by Principal Axis Factoring, considering the Triangular distribution within intervals.}
		\begin{tabular}{|l||cc|c|}
			\hline 
			\textbf{Variable} & \multicolumn{2}{|c|}{\textbf{Estimated factor loadings} } & \textbf{Communalities} \\ 
			& $f_1$ & $f_2$ & \\ 
			\hline
			January & $ -0.9131 $ & $ -0.4000 $ & 0.9937 \\  
			February & $ -0.9371 $ & $ -0.3338 $ & 0.9895 \\  	
			March & $ -0.9735 $ & $-0.2035 $ & 0.9890 \\
			April & $ -0.9925 $ &  +0.0281  & 0.9858  \\  
			May & $ -0.9777 $ & +0.1641 & 0.9827 \\
			June & $ -0.9317 $ & +0.3476 & 0.9889 \\  
			July & $ -0.8954 $ & +0.4407 & 0.9960 \\  
			August & $ -0.9271 $ & +0.3627 & 0.9910 \\ 			
			September & $ -0.9852 $ & +0.1496 & 0.9931 \\   
			October & $ -0.9932 $ & $ -0.0594 $ & 0.9900 \\  
			November & $ -0.9579 $ & $ -0.2549 $ & 0.9825 \\ 
			December & $ -0.9262 $ & $ -0.3703 $ & 0.9950 \\			
			Precipitation & $ -0.8884 $ & +0.2112 & 0.8338 \\
			\hline
			Eingenvalues & 11.6513 & 1.0598 &  \\  
			\hline
			Cumulative proportion of  &  &  &  \\
			total sample variance explained & 0.8963 & 0.9778 &  \\    
			\hline
		\end{tabular} \label{ResumeUnifMeteo}
	\end{table}
\end{center}

Based on the factor loadings of the model we can conclude that all variables: the temperatures of the 12 months of the year and the annual precipitation are strongly related to the 1st factor. Moreover the temperature variables in the months of January and July are moderatly associated with the 2nd factor.  89.6\% of total variance is explained by the 1st factor and 8.2\% by the 2nd factor, which together represent 97.8\% of total variance. All communalities are high indicating that the two factors retained are suitable for describing the latent relational structure between the temperatures of the 12 months of the year and the annual precipitation.\\

The resulting factor model is,\\
\vspace{0,1cm}
\hspace{3,5cm}$January= -0.9131f_{1}-0.4000f_{2}+\varepsilon_{January}$ \\
\vspace{0,1cm}
\hspace{3,4cm}$February = -0.9371f_{1}-0.3338f_{2}+\varepsilon_{February}$ \\
\vspace{0,1cm}
\hspace{3,85cm}$March= -0.9735f_{1}-0.2035f_{2}+\varepsilon_{March}$ \\
\vspace{0,1cm}
\hspace{4,1cm}$April= -0.9925f_{1}-0.0281f_{2}+\varepsilon_{April}$ \\
\vspace{0,1cm}
\hspace{4,2cm}$May= -0.9777f_{1}+0.1641f_{2}+\varepsilon_{May}$ \\
\vspace{0,1cm}
\hspace{4,15cm}$June= -0.9317f_{1}+0.3476f_{2}+\varepsilon_{June}$ \\
\vspace{0,1cm}
\hspace{4,25cm}$July= -0.8954f_{1}+0.4407f_{2}+\varepsilon_{July}$  \\
\vspace{0,1cm}
\hspace{3,8cm}$August= -0.9271f_{1}+0.3627f_{2}+\varepsilon_{August}$ \\
\vspace{0,1cm}
\hspace{3,3cm}$September= -0.9852f_{1}+0.1496f_{2}+\varepsilon_{September}$ \\
\vspace{0,1cm}
\hspace{3,7cm}$October = -0.9932f_{1}-0.0594f_{2}+\varepsilon_{October}$ \\
\vspace{0,1cm}
\hspace{3,3cm}$November= -0.9579f_{1}-0.2549f_{2}+\varepsilon_{November}$ \\
\vspace{0,1cm}
\hspace{3,35cm}$December= -0.9262f_{1}-0.3703f_{2}+\varepsilon_{December}$ \\
\vspace{0,1cm}
\hspace{2,8cm}$Precipitation= -0.8884f_{1}+0.2112f_{2}+\varepsilon_{Precipitation}$ 

\begin{flushleft}
	or,
\end{flushleft}
\vspace{0,1cm}
\hspace{2,85cm}$\Psi_{January}^{-1}(t)= -0.9131\Psi_{f_{1}}^{-1}(1-t)-0.4000\Psi_{f_{2}}^{-1}(1-t)+\Psi_{\varepsilon_{January}}^{-1}(t)$ \\
\vspace{0,1cm}
\hspace{2,6cm}$\Psi_{February}^{-1}(t)= -0.9371\Psi_{f_{1}}^{-1}(1-t)-0.3338\Psi_{f_{2}}^{-1}(1-t)+\Psi_{\varepsilon_{February}}^{-1}(t)$ \\
\vspace{0,1cm}
\hspace{2.95cm}$\Psi_{March}^{-1}(t)= -0.9735\Psi_{f_{1}}^{-1}(1-t)-0.2035\Psi_{f_{2}}^{-1}(1-t)+\Psi_{\varepsilon_{March}}^{-1}(t)$ \\
\vspace{0,1cm}
\hspace{3,15cm}$\Psi_{April}^{-1}(t)= -0.9925\Psi_{f_{1}}^{-1}(1-t)-0.0281\Psi_{f_{2}}^{-1}(1-t)+\Psi_{\varepsilon_{April}}^{-1}(t)$ \\
\vspace{0,1cm}
\hspace{3,2cm}$\Psi_{May}^{-1}(t)= -0.9777\Psi_{f_{1}}^{-1}(1-t)+0.1641\Psi_{f_{2}}^{-1}(t)+\Psi_{\varepsilon_{May}}^{-1}(t)$ \\
\vspace{0,1cm}
\hspace{3,2cm}$\Psi_{June}^{-1}(t)= -0.9317\Psi_{f_{1}}^{-1}(1-t)+0.3476\Psi_{f_{2}}^{-1}(t)+\Psi_{\varepsilon_{June}}^{-1}(t)$ \\
\vspace{0,1cm}
\hspace{3,2cm}$\Psi_{July}^{-1}(t)= -0.8954\Psi_{f_{1}}^{-1}(1-t)+0.4407\Psi_{f_{2}}^{-1}(t)+\Psi_{\varepsilon_{July}}^{-1}(t)$ \\
\vspace{0,1cm}
\hspace{2,9cm}$\Psi_{August}^{-1}(t)= -0.9271\Psi_{f_{1}}^{-1}(1-t)+0.3627\Psi_{f_{2}}^{-1}(t)+\Psi_{\varepsilon_{August}}^{-1}(t)$ \\
\vspace{0,1cm}
\hspace{2,5cm}$\Psi_{September}^{-1}(t)= -0.9852\Psi_{f_{1}}^{-1}(1-t)+0.1496\Psi_{f_{2}}^{-1}(t)+\Psi_{\varepsilon_{September}}^{-1}(t)$ \\
\vspace{0,1cm}
\hspace{2,9cm}$\Psi_{October}^{-1}(t)= -0.9932\Psi_{f_{1}}^{-1}(1-t)-0.0594\Psi_{f_{2}}^{-1}(1-t)+\Psi_{\varepsilon_{October}}^{-1}(t)$ \\
\vspace{0,1cm}
\hspace{2,5cm}$\Psi_{November}^{-1}(t)= -0.9579\Psi_{f_{1}}^{-1}(1-t)-0.2549\Psi_{f_{2}}^{-1}(1-t)+\Psi_{\varepsilon_{November}}^{-1}(t)$ \\
\vspace{0,1cm}
\hspace{2,5cm}$\Psi_{December}^{-1}(t)= -0.9262\Psi_{f_{1}}^{-1}(1-t)-0.3703\Psi_{f_{2}}^{-1}(1-t)+\Psi_{\varepsilon_{December}}^{-1}(t)$ \\
\vspace{0,1cm}
\hspace{2,1cm}$\Psi_{Precipitation}^{-1}(t)= -0.8884\Psi_{f_{1}}^{-1}(1-t)+0.2112\Psi_{f_{2}}^{-1}(t)+\Psi_{\varepsilon_{Precipitation}}^{-1}(t)$
\begin{flushleft}
	with $0 \leq t \leq 1 $.
\end{flushleft}
\vspace{0,2cm}

\begin{figure} [h!] 
	\includegraphics[width=9.5cm, height=9.5cm]{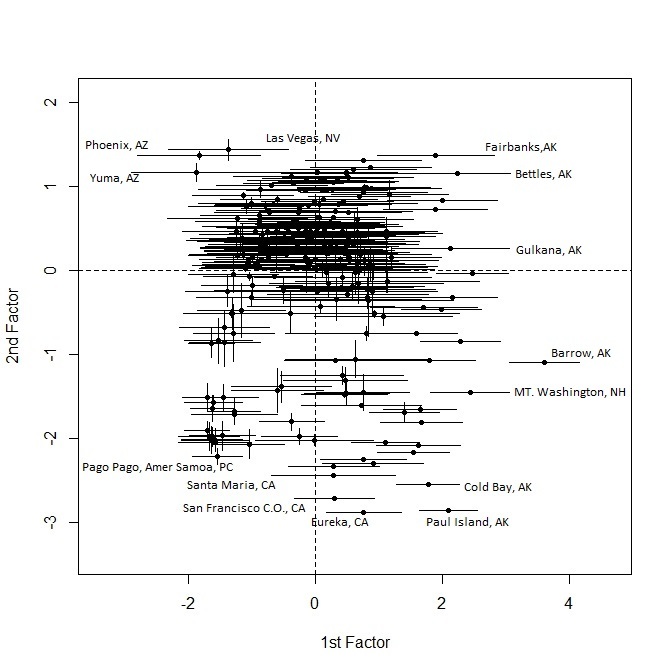}
	\caption{ - Factor scores obtained for the metereological data by the `Anderson-Rubin method'
		and considering the Uniform distribution within intervals.} \label{figARUMeteo}
\end{figure}

The factor scores obtained by the `Anderson-Rubin method' for Meteorological data set are displayed in Table \ref{AndRubUnifMeteo} and represented in Figure \ref{figARUMeteo}.
 
\begin{center}
	\begin{table}
		\caption{ - Factor scores obtained for the metereological data by the `Anderson-Rubin method' and considering the Uniform distribution within intervals.}
		\begin{tabular}{|c||c|c|}
			\hline 
			& Factor 1 & Factor 2 \\ 
			\hline 
			BIRMINGHAM AP, AL & $\left[-1.4742,+0.3535\right]$ & $\left[+0.2627,+0.2763\right]$ \\ 
			\hline 
			HUNTSVILLE, AL & $\left[-1.3915,+0.3445\right]$ & $\left[+0.3439,+0.3532\right]$ \\  
			\hline 
			MOBILE, AL & $[-1.7866,-0.0858]$ & $[+0.0167,+0.0497]$ \\  
			\hline
			MONTGOMERY, AL & $\left[-1.8090,+0.1110\right]$ & $\left[+0.2519,+0.2783\right]$ \\  
			\hline 
			ANCHORAGE, AK & $[+0.9322,+2.2412]$ & $[-0.7550,-0.7504]$ \\  
			\hline    
			&  &   \\
			\vdots & \vdots & \vdots \\
			&  &   \\  
			\hline 
			WAKE ISLAND, PC & $\left[-2.0926,-1.3101\right]$ & $\left[-1.6814,-1.3420\right]$ \\
			\hline 
			YAP, W CAROLINE IS., PC & $\left[-2.0968,-1.0362\right]$ & $\left[-2.1523,-1.8845\right]$ \\  
			\hline 
			SAN JUAN, PR & $\left[-2.0600,-1.1508\right]$ & $\left[-1.6548,-1.4880\right]$ \\  
			\hline 
		\end{tabular} \label{AndRubUnifMeteo}
	\end{table}
\end{center}

Figure \ref{figARUMeteo} shows the 283 cities of the United States of America in the plane defined by the two interval-valued factors. It can be observed that, while the 1st factor distinguishes warm cities with high humidity from cold and dry cities, the 2nd factor basically differentiates cities with larger thermal amplitude from those with short thermal amplitude.\\

\subsubsection{Triangular Distribution}
\label{sec6:1:2}

In this section a Triangular distribution within each observed interval with a randomly generated mode is assumed. 
The data may hence be represented by triplets (min, mode, max) as illustrated in Table \ref{MeteoT}.

\begin{center}
	\begin{table} [h!] 
		\caption{ - Meteorological data set (partial view) described by triplets (min, mode, max).}
		\begin{tabular}{|c||c|c|c|c|} 
	\hline 
	& January & \ldots & December & Precipitation \\ 
	\hline \hline
	BIRMINGHAM AP, AL & $\left(32.3,40.8,52.8\right)$ & \ldots & $\left(35.2,54.2,56.0\right)$ & $\left(41.8,68.8,90.6\right)$ \\ 
	\hline 
	HUNTSVILLE, AL & $\left(30.7,46.1,48.9\right)$ & \ldots & $\left(33.8,43.1,52.4\right)$ & $\left(40.7,67.6,89.4\right)$ \\
	\hline 
	MOBILE, AL & $\left(39.5,59.0,60.7\right)$ & \ldots & $\left(41.6,61.0,62.9\right)$ & $\left(47.8,48.0,91.2\right)$ \\
	\hline 
	MONTGOMERY, AL & $\left(35.5,57.5,57.6\right)$ & \ldots & $\left(37.6,50.4,60.3\right)$ & $\left(43.5,48.2,92.7\right)$ \\
	\hline 
	ANCHORAGE, AK & $\left(9.3,15.1,22.2\right)$ & \ldots & $\left(11.4,13.9,23.7\right)$ & $\left(15.9,49.4,65.3\right)$ \\   
	\hline 
	&  &  &  & \\
	\vdots & \vdots & \vdots & \vdots & \vdots \\
	&  &  &  & \\  
	\hline 
	WAKE ISLAND, PC & $\left(73.1,77.8,82.4\right)$ & \ldots & $\left(74.7,80.9,83.9\right)$ & $\left(76.3,80.3,88.8\right)$ \\  
	\hline 
	YAP, W CAROLINE IS., PC & $\left(73.7,83.6,86.5\right)$ & \ldots & $\left(74.2,86.0,87.0\right)$ & $\left(73.7,83.3,87.7\right)$  \\  
	\hline 
	SAN JUAN, PR & $\left(70.8,73.4,82.4\right)$ & \ldots & $\left(72.1,75.5,83.2\right)$ & $\left(74.0,81.7,87.8\right)$ \\  
	\hline 
\end{tabular} \label{MeteoT}
	\end{table}
\end{center}

Applying the covariance definition $ Cov_3 $ as in formula (\ref{Cov3T_2}) we obtain the following correlation matrix: 

\begin{footnotesize}
	\hspace{1.1cm} Jan 	\hspace{0.1cm} Feb \hspace{0.13cm} Mar \hspace{0.2cm} Apr \hspace{0.2cm} May \hspace{0.2cm} Jun \hspace{0.25cm} Jul \hspace{0.25cm} Aug \hspace{0.2cm} Sept \hspace{0.25cm} Oct \hspace{0.25cm} Nov \hspace{0.2cm} Dec \hspace{0.2cm} Prec
	\begin{equation*}
	\mathbf{R}=\left[
	\begin{array}{ccccccccccccc}
	1 & 0.9815 & 0.9531 & 0.8915 & 0.7978 & 0.6760 & 0.6115 & 0.6757 & 0.8195 & 0.9119 & 0.9643 & 0.9869 & 0.7284 \\
	& 1 & 0.9720 & 0.9212 & 0.8338 & 0.7230 & 0.6676 & 0.7289 & 0.8558 & 0.9246 & 0.9591 & 0.9779 & 0.7506\\
	& & 1 & 0.9582 & 0.8946 & 0.8063 & 0.7485 & 0.8057 & 0.9035 & 0.9526 & 0.9623 & 0.9595 & 0.7945 \\
	& & & 1 & 0.9459 & 0.8860 & 0.8383 & 0.8727 & 0.9367 & 0.9563 & 0.9332 & 0.9087 & 0.8292 \\
	& & & & 1 & 0.9400 & 0.8972 & 0.9114 & 0.9452 & 0.9234 & 0.8689 & 0.8225 & 0.8355 \\
	& & & & & 1 & 0.9396 & 0.9443 & 0.9268 & 0.8641 & 0.7737 & 0.7079 & 0.8152 \\
	& & & & & & 1 & 0.9427 & 0.8990 & 0.8160 & 0.7119 & 0.6425 & 0.8018 \\
	& & & & & & & 1 & 0.9228 & 0.8595 & 0.7652 & 0.7023 & 0.8265 \\
	& & & & & & & & 1 & 0.9402 & 0.8877 & 0.8430 & 0.8584 \\
	& & & & & & & & & 1 & 0.9541 & 0.9268 & 0.8327 \\
	& & & & & & & & & & 1 & 0.9768 & 0.7864 \\
	& & & & & & & & & & & 1 & 0.7440 \\
	& & & & & & & & & & & & 1 \\
	\end{array}
	\right]
	\end{equation*}
\end{footnotesize}

Both Principal Component and Principal Axis Factoring of this matrix leads to the extraction of two factors, with values of estimated factor loadings in interval-valued factors, eigenvalues and communalities very similar. For this reason we only indicate those values for the Principal Axis Factoring method in Table \ref{ResumeTriangMeteo}. 

\begin{center}
	\begin{table} [h!]
		\caption{ - Summary of factor analysis, obtained by Principal Axis Factoring, considering the Triangular distribution within intervals.}
		\begin{tabular}{|l||cc|c|}
			\hline 
			\textbf{Variable} & \multicolumn{2}{|c|}{\textbf{Estimated factor loadings} } & \textbf{Communalities} \\ 
			& $f_1$ & $f_2$ & \\ 
			\hline
			January & $ -0.9105 $ & $ -0.3979 $ & 0.9874 \\  
			February & $ -0.9337 $ & $ -0.3185 $ & 0.9732 \\  	
			March & $ -0.9671 $ & $-0.1871 $ & 0.9703 \\
			April & $ -0.9790 $ &  $-0.0104 $  & 0.9586  \\  
			May & $ -0.9551 $ & +0.1730 & 0.9421 \\
			June & $ -0.9042 $ & +0.3668 & 0.9521 \\  
			July & $ -0.8624 $ & +0.4429 & 0.9399 \\  
			August & $ -0.8993 $ & +0.3630 & 0.9404 \\ 			
			September & $ -0.9663 $ & +0.1488 & 0.9558 \\   
			October & $ -0.9780 $ & $ -0.0550 $ & 0.9960 \\  
			November & $ -0.9533 $ & $ -0.2430 $ & 0.9679 \\ 
			December & $ -0.9270 $ & $ -0.3577 $ & 0.9872 \\			
			Precipitation & $ -0.8556 $ & +0.1389 & 0.7512 \\
			\hline
			Eingenvalues & 11.2671 & 1.0188 &  \\  
			\hline
			Cumulative proportion of  &  &  &  \\
			total sample variance explained & 0.8667 & 0.9451 &  \\    
			\hline
		\end{tabular} \label{ResumeTriangMeteo}
	\end{table}
\end{center}

Based on the factor loadings of the model we can conclude that all variables the temperatures on the 12 months of the year and the annual precipitation are strongly related to the 1st factor. The temperature variables in the months of December and July are moderatly associated with the 2nd factor. 86.7\% of total variance is explained by the 1st factor and only 7.8\% by the 2nd factor, which together represent 94.5\% of total variance. All communalities are high indicating that the two factors retained are suitable for describing the latent relational structure between the temperatures of the 12 months of the year and the annual precipitation.\\

The resulting factor model written as,\\
\vspace{0,1cm}
\hspace{3,5cm}$January= -0.9105f_{1}-0.3979f_{2}+\varepsilon_{January}$ \\
\vspace{0,1cm}
\hspace{3,4cm}$February = -0.9337f_{1}-0.3185f_{2}+\varepsilon_{February}$ \\
\vspace{0,1cm}
\hspace{3,85cm}$March= -0.9671f_{1}-0.1871f_{2}+\varepsilon_{March}$ \\
\vspace{0,1cm}
\hspace{4,1cm}$April= -0.9790f_{1}-0.0104f_{2}+\varepsilon_{April}$ \\
\vspace{0,1cm}
\hspace{4,2cm}$May= -0.9551f_{1}+0.1730f_{2}+\varepsilon_{May}$ \\
\vspace{0,1cm}
\hspace{4,15cm}$June= -0.9042f_{1}+0.3668f_{2}+\varepsilon_{June}$ \\
\vspace{0,1cm}
\hspace{4,25cm}$July= -0.8624f_{1}+0.4429f_{2}+\varepsilon_{July}$  \\
\vspace{0,1cm}
\hspace{3,8cm}$August= -0.8993f_{1}+0.3630f_{2}+\varepsilon_{August}$ \\
\vspace{0,1cm}
\hspace{3,3cm}$September= -0.9663f_{1}+0.1488f_{2}+\varepsilon_{September}$ \\
\vspace{0,1cm}
\hspace{3,7cm}$October = -0.9780f_{1}-0.0550f_{2}+\varepsilon_{October}$ \\
\vspace{0,1cm}
\hspace{3,3cm}$November= -0.9533f_{1}-0.2430f_{2}+\varepsilon_{November}$ \\
\vspace{0,1cm}
\hspace{3,35cm}$December= -0.9270f_{1}-0.3577f_{2}+\varepsilon_{December}$ \\
\vspace{0,1cm}
\hspace{2,8cm}$Precipitation= -0.8556f_{1}+0.1389f_{2}+\varepsilon_{Precipitation}$ 

\begin{flushleft}
	or, using quantile functions, \vspace{0,1cm}
\end{flushleft}
\vspace{0,1cm}
\hspace{2,85cm}$\Psi_{January}^{-1}(t)= -0.9105\Psi_{f_{1}}^{-1}(1-t)-0.3979\Psi_{f_{2}}^{-1}(1-t)+\Psi_{\varepsilon_{January}}^{-1}(t)$ \\
\vspace{0,1cm}
\hspace{2,6cm}$\Psi_{February}^{-1}(t)= -0.9337\Psi_{f_{1}}^{-1}(1-t)-0.3185\Psi_{f_{2}}^{-1}(1-t)+\Psi_{\varepsilon_{February}}^{-1}(t)$ \\
\vspace{0,1cm}
\hspace{2.95cm}$\Psi_{March}^{-1}(t)= -0.9671\Psi_{f_{1}}^{-1}(1-t)-0.1871\Psi_{f_{2}}^{-1}(1-t)+\Psi_{\varepsilon_{March}}^{-1}(t)$ \\
\vspace{0,1cm}
\hspace{3,15cm}$\Psi_{April}^{-1}(t)= -0.9790\Psi_{f_{1}}^{-1}(1-t)-0.0104\Psi_{f_{2}}^{-1}(1-t)+\Psi_{\varepsilon_{April}}^{-1}(t)$ \\
\vspace{0,1cm}
\hspace{3,2cm}$\Psi_{May}^{-1}(t)= -0.9551\Psi_{f_{1}}^{-1}(1-t)+0.1730\Psi_{f_{2}}^{-1}(t)+\Psi_{\varepsilon_{May}}^{-1}(t)$ \\
\vspace{0,1cm}
\hspace{3,2cm}$\Psi_{June}^{-1}(t)= -0.9042\Psi_{f_{1}}^{-1}(1-t)+0.3668\Psi_{f_{2}}^{-1}(t)+\Psi_{\varepsilon_{June}}^{-1}(t)$ \\
\vspace{0,1cm}
\hspace{3,2cm}$\Psi_{July}^{-1}(t)= -0.8624\Psi_{f_{1}}^{-1}(1-t)+0.4429\Psi_{f_{2}}^{-1}(t)+\Psi_{\varepsilon_{July}}^{-1}(t)$ \\
\vspace{0,1cm}
\hspace{2,9cm}$\Psi_{August}^{-1}(t)= -0.8993\Psi_{f_{1}}^{-1}(1-t)+0.3630\Psi_{f_{2}}^{-1}(t)+\Psi_{\varepsilon_{August}}^{-1}(t)$ \\
\vspace{0,1cm}
\hspace{2,5cm}$\Psi_{September}^{-1}(t)= -0.9663\Psi_{f_{1}}^{-1}(1-t)+0.1488\Psi_{f_{2}}^{-1}(t)+\Psi_{\varepsilon_{September}}^{-1}(t)$ \\
\vspace{0,1cm}
\hspace{2,9cm}$\Psi_{October}^{-1}(t)= -0.9780\Psi_{f_{1}}^{-1}(1-t)-0.0550\Psi_{f_{2}}^{-1}(1-t)+\Psi_{\varepsilon_{October}}^{-1}(t)$ \\
\vspace{0,1cm}
\hspace{2,5cm}$\Psi_{November}^{-1}(t)= -0.9533\Psi_{f_{1}}^{-1}(1-t)-0.2430\Psi_{f_{2}}^{-1}(1-t)+\Psi_{\varepsilon_{November}}^{-1}(t)$ \\
\vspace{0,1cm}
\hspace{2,5cm}$\Psi_{December}^{-1}(t)= -0.9270\Psi_{f_{1}}^{-1}(1-t)-0.3577\Psi_{f_{2}}^{-1}(1-t)+\Psi_{\varepsilon_{December}}^{-1}(t)$ \\
\vspace{0,1cm}
\hspace{2,1cm}$\Psi_{Precipitation}^{-1}(t)= -0.8556\Psi_{f_{1}}^{-1}(1-t)+0.1389\Psi_{f_{2}}^{-1}(t)+\Psi_{\varepsilon_{Precipitation}}^{-1}(t)$
\begin{flushleft}
	with $0 \leq t \leq 1 $.
\end{flushleft}
\vspace{0,2cm}

\begin{figure} [h!] 
	\includegraphics[width=9.5cm, height=9.5cm]{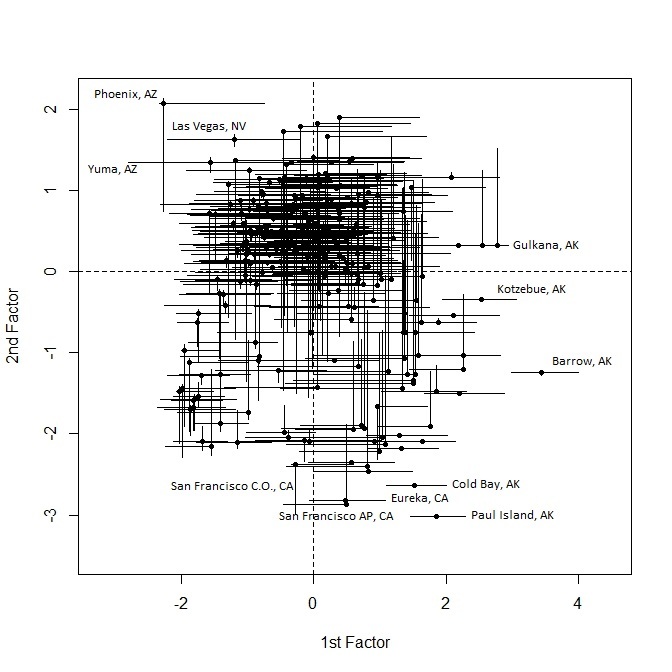}
	\caption{ - Factor scores obtained for the metereological data by the `Anderson-Rubin method'
		and considering the Triangular distribution within intervals.} \label{figARTriMeteo}
\end{figure}

The factor scores obtained by the `Anderson-Rubin method' for Meteorological data set are displayed in Table \ref{AndRubTriMeteo} and represented in Figure \ref{figARTriMeteo}.

\begin{center}
	\begin{table}
		\caption{ - Factor scores obtained for the metereological data by the `Anderson-Rubin method' and considering the Uniform distribution within intervals.}
		\begin{tabular}{|c||c|c|}
			\hline 
			& Factor 1 & Factor 2 \\ 
			\hline 
			BIRMINGHAM AP, AL & $\left(-1.5239,-0.5693,+0.3565\right)$ & $\left( +0.2263,+0.2907,+0.2725\right)$ \\ 
			\hline 
			HUNTSVILLE, AL & $\left(-1.3746,-0.6350,+0.3916\right)$ & $\left(+0.4210,+0.4615,+0.4779\right)$ \\  
			\hline 
			MOBILE, AL & $(-1.7899,-0.9609,-0.0203)$ & $(-0.1760,-0.1071,+0.0522)$ \\  
			\hline
			MONTGOMERY, AL & $\left(-1.7872,-0.7832,+0.2274\right)$ & $\left(-0.0588,-0.0061,+0.1055\right)$ \\  
			\hline 
			ANCHORAGE, AK & $(+0.8832,+1.5452,+2.1446)$ & $(-0.7528,-0.7528,-0.7499)$ \\  
			\hline    
			&  &   \\
			\vdots & \vdots & \vdots \\
			&  &   \\  
			\hline 
			WAKE ISLAND, PC & $\left(-2.0054,-1.9622,-1.4283\right)$ & $\left(-1.9039,-0.9689,-0.9016\right)$ \\
			\hline 
			YAP, W CAROLINE IS., PC & $\left(-2.2681,-1.8660,-1.1843\right)$ & $\left(-2.0128,-1.6979,-1.6445\right)$ \\  
			\hline 
			SAN JUAN, PR & $\left(-2.2222,-1.6949,-1.2837\right)$ & $\left(-1.3940,-1.2853,-1.2500\right)$ \\  
			\hline 
		\end{tabular} \label{AndRubTriMeteo}
	\end{table}
\end{center}

Figure \ref{figARTriMeteo} shows the 283 cities of the United States of America in the plane defined by the two interval-valued factors. The conclusions are exactly the same as those taken earlier, the 1st factor distinguishes warm cities with high humidity from cold and dry cities and the 2nd factor differentiates cities with larger thermal amplitude from those with short thermal amplitude.
We notice however that in this analysis there are less degenerate intervals in the second factor, than in the analysis that assumed an Uniform distribution.

\section{Concluding remarks}
\label{sec:6}

Most of the methodologies developed for Symbolic Data Analysis rely on distribution free approaches. In this paper we addressed the analysis of the dependence structure of interval-valued variables, proposing a factor model for interval data based on quantile function representations. In our proposal, factor extraction is carried on by performing a Principal Component or a Principal Axis Factoring based on the correlation matrix between the observed interval-valued variables. For that purpose we rely on, and extend, appropriate definitions of variance, covariance and correlation, for interval variables under the assumptions of uniform or triangular distributions to model the within variability of each interval.  Factor scores were derived by solving optimization problems, that adapt the Bartllet, and Anderson-Rubin methods for real-valued data.  However, unlike the original problems, in the case of interval data, the resulting optimization problems do not have a closed-formal analytical solution, and need to be solved numerically.  
The research presented in this paper may be extended in several ways.  In the first place, alternative methods of factor extraction can be devised.  One important avenue of research  is the study of parametric methods of factor extraction, based on existing models for interval data, such as those proposed in Brito and Duarte Silva \cite{Brito2012}.  Secondly, factor rotation of interval-valued factors may be adressed. Finally, the basic approach proposed here can be extended to establish factor models for other types of symbolic data such as distributional or histogram data.\\

\vspace{0.4cm}

\textbf{Acknowledgements}\\

This work is financed by  National Funds through the FCT - Funda\c{c}\~{a}o para a Ci\^{e}ncia e Tecnologia (Portuguese Foundation for Science and Technology) as part of projects UID/EEA/50014/2013 and  UID/GES/00731/2013.




\end{document}